\documentclass[%
 reprint,
 amsmath,amssymb,aps,prl,superscriptaddress,nolongbibliography]{revtex4-2}

\usepackage{braket}
\usepackage{graphicx}
\usepackage{dcolumn}
\usepackage{hyperref}
\usepackage{bm}
\usepackage{xcolor}

\begin{document}
\title{Attosecond Optical Orientation}

\author{Lauren B Drescher}
\email{lauren.drescher@mbi-berlin.de}
\affiliation{Department of Chemistry, University of California, Berkeley, California 94720, USA}
\affiliation{Chemical Sciences Division, Lawrence Berkeley National Laboratory, Berkeley, California 94720, USA}
\affiliation{Max-Born-Institut, Max-Born-Str. 2A, 12489, Berlin, Germany}
\author{Nicola Mayer}
\affiliation{Attosecond Quantum Physics Laboratory, Department of Physics, King's College London, Strand, London, WC2R 2LS, United Kingdom}
\affiliation{Max-Born-Institut, Max-Born-Str. 2A, 12489, Berlin, Germany}

\author{Kylie Gannan}
\affiliation{Department of Chemistry, University of California, Berkeley, California 94720, USA}
\author{Jonah R Adelman}
\affiliation{Department of Chemistry, University of California, Berkeley, California 94720, USA}
\author{Stephen R Leone}
\affiliation{Department of Chemistry, University of California, Berkeley, California 94720, USA}
\affiliation{Chemical Sciences Division, Lawrence Berkeley National Laboratory, Berkeley, California 94720, USA}
\affiliation{Department of Physics, University of California, Berkeley, California 94720, USA}

\date{\today}

\begin{abstract}
Circularly polarized light offers opportunities to probe symmetry-dependent properties of matter such as chirality and spin. Circular dichroism measurements typically require further intrinsic or extrinsic breaking of symmetry by e.g. enantiomeric excess, orientation, magnetic fields or direction-sensitive detectors. Here we introduce circular-dichroic attosecond transient absorption spectroscopy by leveraging the angular momentum of two circular-polarized pulses, both pump and probe, in an isotropic medium, optically orienting the angular momentum of excited states on an attosecond timescale. We investigate a circular-dichroic measurement of the attosecond transient absorption of He Rydberg states. By limiting the allowed pathways via dipole selection rules for co- and counter-rotating circular polarized NIR and XUV pulses, different spectral reshapings of the XUV transient absorption due to the AC Stark effect are observed. Paired with time-dependent Schrödinger equation calculations, the results show the role of selection and propensity rules and open up new opportunities to study coupling pathways of excited states as well as spin-dependent dynamics in atoms and beyond via attosecond optical orientation.

\end{abstract}

\maketitle

Attosecond spectroscopy promises to follow electron dynamics in atoms, molecules and solids on their natural timescale~\cite{leone2014}. Recent advances in the generation of circularly polarized attosecond light sources extend the opportunities to explore time-dependent dynamics that couple to the spin angular momentum of light.
Most prominently, it enabled the study of time-dependent magnetization and spin dynamics in solids at the attosecond timescale under the influence of external magnetic fields~\cite{willems2015,siegrist2019,geneaux2024}. However, many more spin and angular momentum dependent phenomena of interest exist in the absence of external magnetic fields, such as dimensional and topological~\cite{hasan2010,schaibley2016,bao2021}, chiral~\cite{yang2021,rouxel2022} and orientational effects~\cite{franzen1957,skrotskii1961,meier1984}.

\begin{figure}[htbp]
    \includegraphics[width=.5\textwidth]{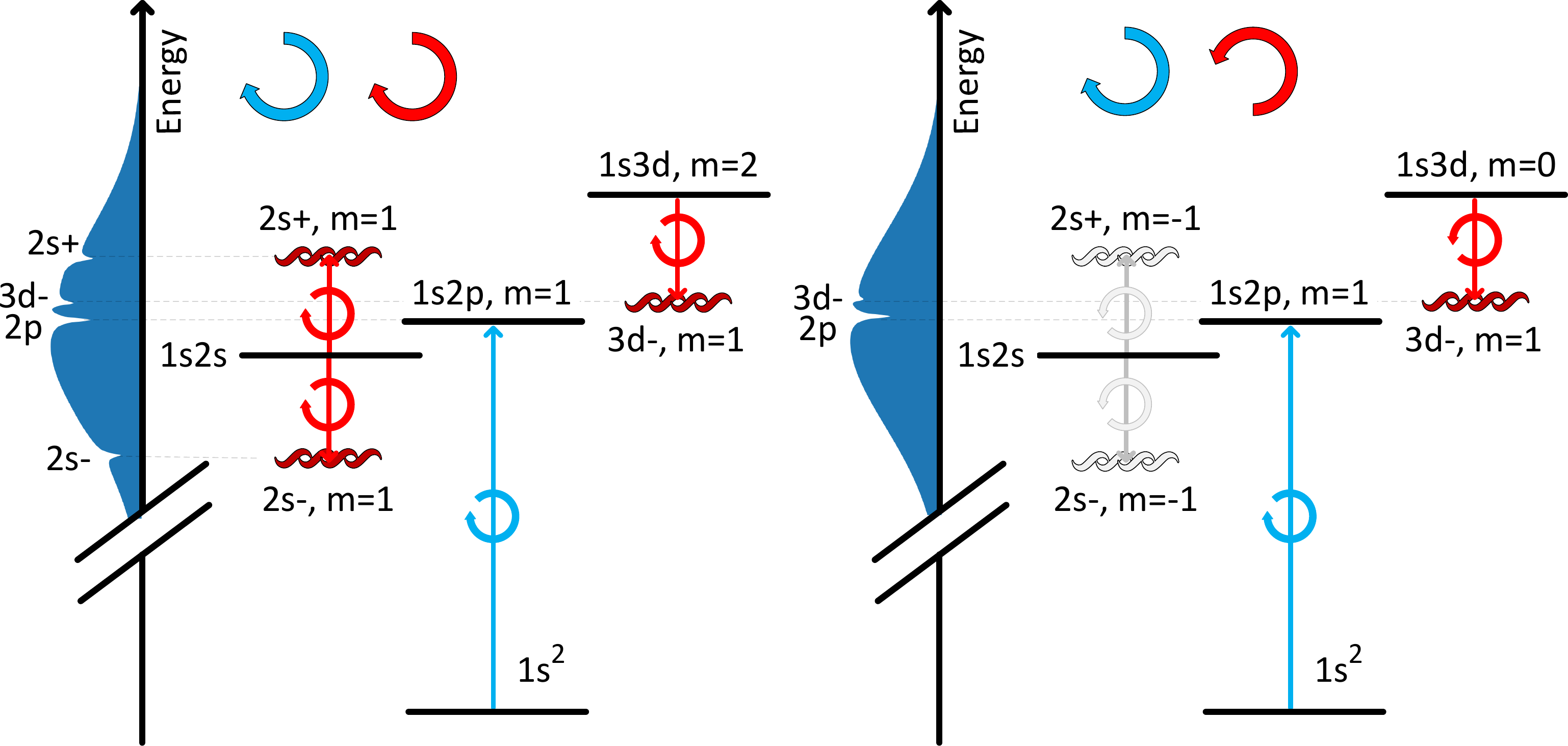}
    \caption{Experimental scheme: A circularly polarized XUV pulse excites the 1s2p($m=1$) level of He. In the presence of a co-rotating NIR pulse (left), coupling to the 1s2s and 1s3d levels leads to the presence of LISs in the XUV spectrum. When the NIR pulse is counter-rotating to the XUV pulse (right), coupling to the 1s2s state is forbidden and the 2s-LISs disappear.  
    }\label{fig:schemes}
\end{figure}

Here, we introduce circular-dichroic attosecond transient absorption spectroscopy (cDATAS) to study $m$ quantum number dependent coupling in isotropic media, without an external magnetic alignment field and by optical orientation~\cite{franzen1957,skrotskii1961,meier1984} with co- and counter-rotating circularly polarized XUV+NIR two-color fields. Leveraging the selection and propensity rules in absorption and emission, we preferentially populate given excited states and show how magnetic sublevel selectivity determines the spectrum of an atom. Through the coupling of orbital and spin angular momentum, this method allows to study electron spin-dynamics~\cite{meier1984} with attosecond time methodology. While we focus on the attosecond optical orientation in the absence of magnetic fields in this work, optical orientation previously has been used to develop sensitive atomic magnetometers~\cite{franzen1957,skrotskii1961}, opening the possibility to bring all-optical quantum sensing to the ultrafast domain~\cite{sutcliffe2024}.
We investigate the principle experimentally on the $m$-sublevel dependent coupling of the He 1s$n$p-Rydberg series, which has been a prototypical system for linearly polarized attosecond transient absorption spectroscopy (ATAS)~\cite{wu2016,Chini:2012aa,reduzzi2015}, together with time-dependent Schrödinger equation (TDSE) calculations and a perturbative model.
Switching the NIR helicity thereby controls the appearance of light induced structures (LISs) in the absorption spectrum of He, depending on its optical orientation (see Fig.~\ref{fig:schemes}).

The presented study is similar in concept to ultrafast and attosecond photoionization spectroscopy studies using circular polarized, as well as co- and counter-rotating pulses in strong-field~\cite{hartung2016,jimenez-galan2018,mayer2022,carlstrom2024} or in multiphoton processes~\cite{eckart2018,desilva2021,han2023,kheifets2024}, however targeting the light-induced coupling between bound states and offering the excellent temporal resolution paired with the spectral resolution that transient all-optical spectroscopy methods offer.

\begin{figure}[htbp]
    \includegraphics{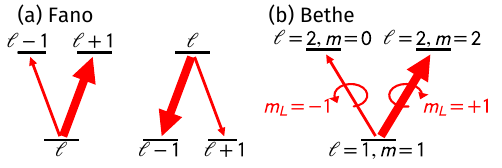}
    \caption{Fano's propensity rule (a) states that $\ell$ tends to increase in absorption (left) and decrease in emission (right), as indicated by arrow size. Bethe's propensity rule (b) indicates that $|m|$ increases (for absorption and emission), as indicated for the transition from a $\ell=1,m=1$ state. Additionally, dipole selection rules requires $\Delta m=m_L$, where $m_L$ is the spin angular momentum of light.
    }\label{fig:propensity}
\end{figure}

In attosecond transient absorption, the XUV pulse induces a time-dependent dipole moment in the medium, which is modified by another pulse, such as a NIR optical pulse. The resulting emission interferes with the incoming XUV pulse, leaving fingerprints of the interaction on the XUV spectrum. In the approximation of a thin gas jet, the ATAS signal is directly related to the atomic response via \cite{wu2016}
$S(\omega)\propto\,2\omega\, \text{Im}\left[\tilde{\mathbf{d}}(\omega)\cdot\tilde{\mathbf{E}}^*_{XUV}(\omega)\right]$,
where $\tilde{\mathbf{d}}(\omega)$ and $\tilde{\mathbf{E}}(\omega)$ are the Fourier transforms of the time-dependent dipole moment of the medium and the XUV electric field, respectively. Adding the delay-dependent NIR-field allows to prepare specific excited states, dress the medium or perturb the XUV induced dipole moment, depending on the arrival of the NIR pulse before, during or after the XUV pulse, respectively. Changing the time delay $\tau$ between the two pulses, the underlying ultrafast dynamics can be studied with attosecond precision.
In an isotropic medium like a gas jet of atoms, the accessible transitions are determined by the polarization of the incoming fields via the selection rules imposed on the angular momentum $\ell$ and magnetic quantum number $m$ of the electronic states involved. In the dipole approximation, these are $\Delta\ell=\pm1$ and $\Delta m=m_L$, where $m_L$ is the spin angular momentum of light. The strength of a given transition is determined by its dipole matrix element $\braket{\Psi_{n\ell m}|\mathbf{d}|\Psi_{n'\ell'm'}}$, and well-known propensity rules govern the relative strength of the different transitions allowed by the selection rules. In particular, Fano's propensity rule dictates that the absorption (emission) of photons tends to increase (decrease) the angular momentum $\ell$ (see Fig.~\ref{fig:propensity}(a))~\cite{Fano:1985aa}. It derives from the radial part of the dipole integral $\int_0^{\infty}dr\Psi^*_{n\ell}(r)r\Psi_{n'\ell'}(r)$, or more precisely from the contribution arising from the centrifugal term $\ell(\ell+1)/2r^2$. Another propensity rule, stated by Bethe \cite{Bethe:1997aa}, dictates that both absorption or emission of a photon tends to increase the absolute value of the magnetic quantum number ($|m|$), i.e. transition from an $m=-1$ to $m=-2$ sublevel is preferred over $m=0$, or transition from an $m=2$ to an $m=3$ is preferred over $m=1$, for the same $\ell$ final state (see Fig.~\ref{fig:propensity}(b)). This rule derives from the angular part of the dipole integral, known as the Gaunt coefficient. These propensity rules arise in strong-field processes like bicircular high-harmonic generation~\cite{jimenez-galan2018} and have been observed at play in photoelectron spectroscopy experiments \cite{busto2019,han2023}. To date, they remain unstudied in ATAS experiments. Our work extends upon previous studies on the linear dichroism in attosecond transient absorption, i.e. the dependence of the dipole coupling on parallel or orthogonal polarization of XUV and NIR pulses~\cite{reduzzi2015} and extends the control over the angular momentum quantum number $\ell$ in the linear dichroic case to the magnetic quantum number $m$ in circular dichroism.

\begin{figure}[htbp]
    \includegraphics{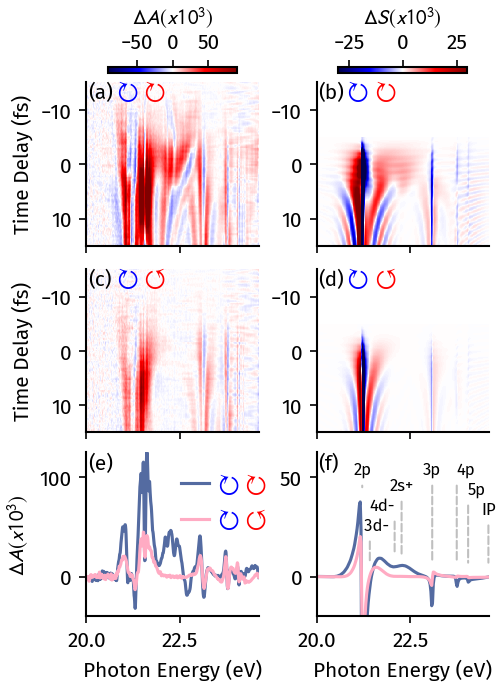}
    \caption{(a,c) Experimental and (b,d) calculated transient absorption spectra of the He 1s$n$p Rydberg series for (a/b) co- and (c/d) counter-rotating circularly polarized XUV+NIR pulses. Taking the transient spectra at $\tau=1.3$\,fs, the experimental results (e), as well as the calculations using TDSE (f) show a strong enhancement of the change of absorbance ($\Delta A$) in the co- over the counter-rotating case due to the increased coupling of bright states of the $n$p series to dark states of the $n$s and $n$d series. The light-induced states (LISs) associated with the 2s+, 3d- and 4d-states (around 22\,eV) are strongly suppressed in the counter-rotating configuration.
    }\label{fig:exp_map}
\end{figure}

To perform the experiment, a typical attosecond transient absorption spectroscopy setup~\cite{geneaux2024,deroulet2024} is modified to perform the circular-dichroic measurements: A four-mirror phase-retarderis inserted in the XUV beam path, to transform the linearly polarized XUV radiation from High Harmonic Generation into a highly elliptical polarization state and motorized quarter-waveplate in the NIR arm is used to switch the NIR helicity. A detailed description can be found in Appendix A.
The resulting transient absorption spectra for co- and counter-rotating configurations of the NIR and XUV pulses are shown in Fig.~\ref{fig:exp_map}(a) and (c), respectively.

In the co-rotating case, the spectra show strong similarity to previous measurements on the attosecond transient absorption of He Rydberg states for co-linear polarized pulses~\cite{Chini:2012aa,reduzzi2015,wu2016}. Around the bright excited states of the 1s$n$p series converging towards the ionization potential at 24.59\,eV, modified lineshapes are visible  during and after temporal overlap of the XUV and NIR pulses due to the light-induced phase that accumulates from NIR-driven coupling to nearby dark states, as well as hyperbolically converging lines towards larger delays due to the truncation of the exponential decay of the time-dependent dipole moment due to NIR driven population transfer (perturbed free induction decay). In between the bright states, new resonant structures appear during overlap with the NIR pulse due to the NIR dressing of the atom, which allows for the population of one-photon dark states of the 1s$n$s and 1s$n$d series. Population is transferred to these states by additional absorption or emission of a NIR photon. Consequently, their resonance appears at plus or minus one NIR photon energy, respectively, below or above their field-free energies. We label these LISs according to their state configuration and whether their excitation followed the absorption (+) or emission (-) of a NIR photon.
Such LISs can be prominently found in Fig.~\ref{fig:exp_map}(a) between the 1s2p Rydberg state (21.22\,eV) and 1s3p Rydberg state (23.09\,eV), e.g. the 3d- ($\sim$21.4\,eV), 4d- ($\sim$22.1\,eV), 2s+ ($\sim$22.3\,eV) LISd, or 3s- LIS ($\sim$21.3\,eV). Due to the broad-band nature of the NIR and the decreasing spacing between Rydberg states, assignment of the individual LISs is increasingly difficult approaching the ionization potential.

The strength of many of these features appears increased compared to the case of a linearly polarized NIR pulse with the same XUV ellipticity (not shown). More strikingly, in case of a counter-rotating NIR pulse, these features appear strongly reduced if not absent as can be seen in Fig.~\ref{fig:exp_map}(c):
while the effect of the light-induced phase seems substantially reduced around the bright state resonant energies, the LISs of the 1s$n$s series are not observed and the LIS of the 1s$n$d series is heavily suppressed.

\begin{figure}[htbp]
    \includegraphics{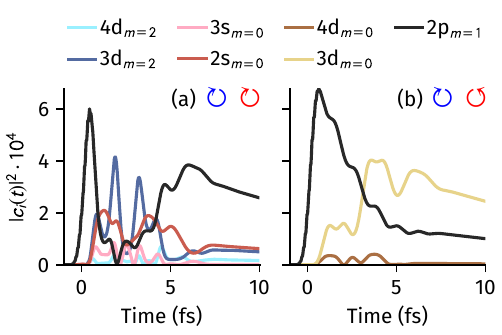}
    \caption{(a,b) Time-dependent populations of states from TDSE. The $n$p$_{m=1}$ states are populated by the circularly-polarized XUV pulse at 0\,fs and start to decay. Population is transferred between neighboring (dark) states by the NIR pulse (centered at 1.3\,fs), depending on the selection and propensity rules for the (a) co- and (b) counter-rotating NIR helicity. Only states that are populated are shown.
    }\label{fig:theo_pop}
\end{figure}

To get further physical insight into the observations, time-dependent Schrödinger equation (TDSE) calculations are employed.
We use a restricted few-level model to solve the TDSE $\text{i}\partial_t|\Psi(t)\rangle=H(t)|\Psi(t)\rangle$, as explained in detail in Appendix B. In particular, we include the 1s$n$s series up to the $n=6$ state, the 1s$n$p series up to the $n=5$ state and the 1s$n$d series up to the $n=5$ state, as well as the ground state 1s$^2$. 
The NIR and XUV pulse are approximated by $\sin^2$ envelope pulses, respectively 15 fs and 1.74 fs long, with central wavelengths of $\lambda_{NIR}=780$\,nm and $\lambda_{XUV}=57$\,nm and intensities $I_{NIR}=3.51\cdot10^{12}$\,W/cm$^2$ and $I_{XUV}=0.87\cdot10^{12}$\,W/cm$^2$. 
The results are shown in Fig.~\ref{fig:exp_map}(d,f) as the single-atom response ($\Delta S$), i.e. without propagation effects and scaling to the experimental absorbance ($\Delta A)$. 
The color maps show the ATAS spectrum between -5\,fs and 15\,fs time delay between the XUV and the NIR pulse for co-rotating (Fig.~\ref{fig:exp_map}(b)) and counter-rotating (Fig.~\ref{fig:exp_map}(d)) pulses, where the XUV is set to be perfectly left circularly polarized (LCP) in the calculation. 
It is clear that a LIS feature at $\simeq 22.5$ eV beating at a $2\omega_\textrm{NIR}$ frequency is observed only in the co-rotating case. Its energy matches the 2s+ LIS (see Fig.~\ref{fig:exp_map}b/d). 
The expected position of a LIS feature associated with the emission of one NIR photon from the 1s4d state (4d-) is also indicated. On top of the 2s+/4d- LISs, the LISs arising from the emission of one NIR photon from the 1s3s and 1s3d states also play a role, as these are almost resonant with the 1s2p line (see assignments in Fig.~\ref{fig:exp_map}(f)). In particular, the 3d- LIS splits the resonance profile when the XUV and NIR pulses overlap, reconnecting with the main 1s2p line after the overlap is over at $\tau\simeq7.5$ fs. $2\omega_{NIR}$ oscillations are seen from this LIS as well, as expected. Note that the 3d- LIS splits the profile of the resonance in the counter-rotating case as well, as seen in Fig.~\ref{fig:exp_map}(d).

The time-dependent population of the excited states from TDSE calculations for the two cases are shown in Fig.~\ref{fig:theo_pop} for a XUV+NIR time delay of $\tau=1.3$\,fs (i.e. when the 2s+ LIS is clearly present for the co-rotating case). It is clear that the NIR has the effect of shifting population off-resonantly between the $1s2p_{m=1}$ state and the $1s2s_{m=0}$ and $1s3d_{m=2}$ state. The population that leaves the $1s2p_{m=1}$ state oscillates between these two Rydberg series, with larger final population on either state depending on the relative delay between the XUV and NIR pulses, as seen also in the anti-phase beatings of their corresponding LIS features in the ATAS spectrum of Fig.~\ref{fig:exp_map}(b). Minor population is also moved to the $1s3s_{m=0}$ and $1s4d_{m=2}$ states.

In the counter-rotating case, the $m=2$ states are no longer accessible, and population transfer occurs mainly between the $1s2p_{m=1}$ and the $1s3d_{m=0}$ state. Notably, no population is transferred to the $1s2s_{m=0}$ state, as we would expect from the selection rules, since emission of a counter-rotating photon from the $1s2p_{m=1}$ state would lead to a final spin angular momentum $m>0$. Correspondingly, no LIS feature at $22.5$ eV is observed in the ATAS map of Fig.~\ref{fig:exp_map}(c,d).

Comparing the spectra obtained from TDSE calculations to the experimentally observed spectra, we note that while the general trends of co- vs counter-rotating pulses are well reproduced by the TDSE calculation, several differences are observed, especially in the specific lineshapes close to resonance. This is attributed to uncompressed satellite pulses in the NIR from higher-order dispersion as well as the relatively high gas pressure used in the experiment ($\sim$80\,mbar), which is known to cause spectral reshaping (resonant-propagation~\cite{liao2015}) and novel spectral structures to emerge in the vicinity of the resonances~\cite{pfeiffer2013}, as can for example be observed around the 2p resonance in Fig~\ref{fig:exp_map}(a/e).

\begin{figure}
    \includegraphics{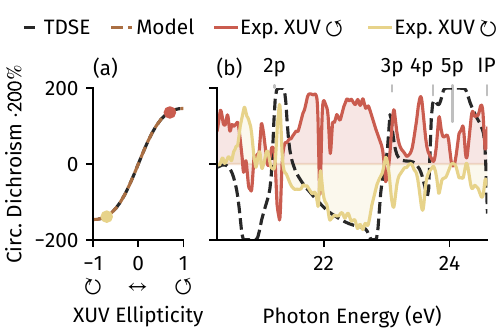}%
    \caption{(a) Circular dichroism of the 2s+ LIS as a function of XUV ellipticity for the TDSE calculation (black), perturbative model (dashed orange) and experimental results for LCP (yellow) and RCP (red). (b) Experimental and TDSE dichroism spectrum at temporal overlap of XUV and NIR pulses. When XUV helicity is flipped, the dichroism is inverted. The TDSE spectrum is shown for LCP XUV.\label{fig:exp_lineout}}
\end{figure}

Since the 2s+ LIS feature is the most striking dichroic signal we observe in both experiment and theory, we focus on its circular dichroism. Fig.~\ref{fig:exp_lineout}(a) shows the dichroic ATAS signal integrated around this LIS from 22.1 to 22.5\,eV, where we show the normalized circular dichroism $\text{CD}(\omega)=\frac{S_{\circlearrowright}(\omega)-S_{\circlearrowleft}(\omega)}{|S_{\circlearrowright}(\omega)|+|S_{\circlearrowleft}(\omega)|}\cdot200\%$, where $S_{\circlearrowright}(\omega)$ ($S_{\circlearrowleft}(\omega)$) corresponds to the ATAS signal for LCP (RCP) NIR pulse. In particular, we study the CD dependence on the XUV ellipticity $\epsilon_{XUV}$, where $\epsilon_{XUV}=\pm1$ corresponds to LCP/RCP, respectively. The results are compared to a perturbative model (see Appendix C), which predicts that $\text{CD}(\omega_\textrm{LIS})\propto\frac{2\epsilon_X}{1+\epsilon_X^2}$. As can be seen in Fig.~\ref{fig:exp_lineout}(a), the results of the TDSE calculation follows the perturbative result closely. The circular dichroism is maximized for circularly polarized XUV and is zero for linear XUV, as one would expect. Moreover, the CD signal for opposite XUV ellipticities is reversed.
This confirms that the observed dichroism originates in the different helicities of the XUV and NIR pulses: The experiments were repeated with the XUV polarizer set to the opposite helicity, which as expected reverses the role of co- and counter-rotating helicity of the NIR. Fig.~\ref{fig:exp_lineout}(b) shows the circular dichroism (averaged over 2\,fs around $\tau=1.3$\,fs) for LCP and RCP XUV pulses, respectively. Due to the flip of the XUV helicity, the dichroism appears inverted. Small differences in the experimental signal are attributed to a drift in the measurement conditions between experiments. Compared to the circular dichroism from TDSE calculations (dashed line in Fig.~\ref{fig:exp_lineout}(b)), the experimental signal agrees well with the calculation especially in the region between the 2p and 3p resonances. A comparison of the integrated signal around the 2s+ LIS (dots in Fig.~\ref{fig:exp_lineout}(a)) shows good agreement. At higher energies, the calculated signal shows an inverted sign, possibly due to the influence of the continuum, which is not included in the few-level approximation. Smaller deviations are attributed to the limited circularity of the XUV pulse in the experiment, as well as macroscopic propagation and pulse structure effects.

Our results show that the selective excitation of $m$ quantum number states due to the total orbital momentum coupling of circular polarized light allows to optically orient excited states in isotropic media, such as gases and centrosymmetric crystals, by investigating the spectral response of the XUV transient absorption due to co- and counter-rotating NIR pulses. The spectra thereby give insight into the electron configuration of bright excited states and their (dark state) neighbors. The method allows to explore $\Lambda$- versus $V$-type couplings, similar to four-wave-mixing~\cite{gaynor2021,gaynor2023}, however here due to their different selection rules of $m$-quantum numbers. A configuration that wasn't explored here additionally allows to isolate ladder-type couplings between states: A ladder-type coupling e.g. $\ell\to(\ell+1)\to\ell$ is not allowed for circular polarized light, while being allowed for linear polarized light.
The CD behavior is not specific to this He Rydberg series but is a general result of dipole selection and propensity rules~\cite{busto2019}. The control over the $m$ quantum number due to the circular-dichroism in He affects mainly the orbital angular momentum, due to the very weak spin-orbit coupling of He, making it an ideal system to demonstrate the principle of the CD method. However, stronger spin-orbit coupling in different systems that depart from LS-coupling will allow the XUV CD method to examine and control spin-specific excited states.
As such, we have applied the cDATAS method to a series of more complex systems, in particular the Ar and Ne autoionization resonances and solid-state LiF core-exciton resonances, which will be discussed in detail in future publications.

The results shown here are symmetric under simultaneous change of helicity of the two-color field. This property arises from inversion symmetry and is not expected to hold for parity-broken systems~\cite{drescher2020}, which promises novel insights for e.g. chiral molecules and non-centrosymmetric crystals~\cite{rouxel2022}.
While we have focused in this letter on the circular dichroism of the NIR-induced coupling in XUV-excited states, the method is also applicable to NIR- or optically-excited states. This will allow to study the evolution and spin dynamics of optically excited states without requiring an external magnetic field~\cite{meier1984}, with the extreme temporal resolution, high spectral resolution and core-level specificity of XUV transient absorption spectroscopy. One area of high interest for this method will be materials for valleytronics due to the locking of the valley-specific (pseudo-)spin~\cite{geondzhian2022,malakhov2024}.

\section*{Acknowledgments}

Investigations were supported by the U.S. Air Force Office of Scientific Research, Grant Nos FA9550-24-1-0184 and FA9550-19-1-0314. LBD acknowledges the European Union's Horizon research and innovation programme under the Marie Sklodowska-Curie grant agreement No 101066334 — SR-XTRS-2DLayMat. NM acknowledges helpful discussions with Felipe Morales. This work was funded by UK Research and Innovation (UKRI) under the UK government’s Horizon Europe funding guarantee [grant number EP/Z001390/1]. JRA acknowledges the NSF GRFP under grant No DGE 2146752 and the U.S. Department of Energy, Office of Science, Basic Energy Science (BES), Materials Sciences and Engineering Division under contract DE-AC02-05CH11231 within the Fundamentals of Semiconductor Nanowires Program (KCPY23) for personnel support.

\section*{Appendix A: Description of the experimental setup}
NIR pulses from a commercial regenerative Ti:Sa amplifier are spectrally broadened through self-phase modulation in a gas-filled stretched hollow-core fiber and compressed in time by an array of chirped mirrors and dispersive material, to a duration of less than 4\,fs. The few-cycle pulse is then split in two parts at a beam splitter, with most of the pulse power being focused into a vacuum beamline into an Kr gas-flushed cell to generate XUV attosecond pulses through the process of High Harmonic Generation (HHG). The generated broadband XUV pulses are filtered from the residual driving pulse by passage through a thin Al foil and refocused by a grazing incidence toroidal mirror. Afterwards the linearly polarized XUV beam passes through a four-mirror phase-retarder (prototype of Ultrafast Innovation Aurora). The four-fold reflection by protection-coated Mo-plates under a selectable angle leads to a phase delay between the s- and p-components of the XUV polarization and therefore produces highly elliptical polarization ($\epsilon_\textrm{XUV}\sim\pm0.7$)~\cite{vonkorffschmising2017}. After passage through the polarizer, the beam focuses into the absorption cell flushed with the He target gas. The previously split-off weaker part of the few-cycle pulse travels alongside the vacuum beamline on an optical table, where a piezo-controlled delay stage allows to vary the pulse travel time to control the time-delay in the experiment. After reflection by a focusing mirror the NIR beam is co-linearly recombined with the XUV beam inside the vacuum beamline by a hole-mirror, but not before transmission through a achromatic quarter-waveplate (B. Halle RAC 4.4) mounted in a direct-drive rotation mount, allowing to rapidly modify the polarization state of the NIR pulses with varying helicity.
After both beams are transmitted through the He absorption cell, a second Al foil filters the residual NIR pulse and the XUV pulse is spectrally dispersed on a flat-field grating and the spectra are imaged by a XUV-sensitized CCD camera.
Additionally, a mechanical shutter in the NIR beam path allows to record XUV spectra in the presence of the NIR pulses ($I_\textrm{on}(\omega)$) and with the NIR pulses blocked ($I_\textrm{off}(\omega)$) in rapid succession to minimize artifacts from slow variations in the XUV spectra.
This allows the delay-dependent change in absorbance $\Delta A$ due to the pump pulse to be obtained as $\Delta A(\omega,\tau) = -\log_{10}[I_\textrm{on}(\omega,\tau)/I_\textrm{off}(\omega)]$.

In the experiment, the time delay between NIR and XUV pulses was varied between -15\,fs to 18\,fs, with 200 steps (0.166\,fs steps), where the NIR pulse arrives after the XUV pulse for positive delays. A variable iris in the pump arm allows control of the NIR intensity and is set to an estimated peak intensity of $2\cdot10^{12}$\,W/cm$^2$. The instrument response function (IRF) was determined to be 5.8$\pm$0.6\,fs through NIR-induced modification of the Ne 2s$^1$2p$^6$3p$^1$ autoionizing resonance. To minimize artifacts due to changing beam properties, co- and counter-rotating spectra were taken in rapid succession by rotating the NIR quarter-waveplate between the +45$^\circ$ and -45$^\circ$ position.

\section*{Appendix B: Effective few-level model}
In the effective model we include only the relevant states and transitions. The TDSE $\text{i}\partial_t|\Psi(t)\rangle=H(t)|\Psi(t)\rangle$ is solved in the length gauge by expanding the wavefunction on the basis of bound states $|\Psi(t)\rangle=c_g(t)|1s^2\rangle+\sum_{n\ell m}c_{n\ell m}(t)|n,\ell,m\rangle$, where $|1s^2\rangle$ is the helium ground state and we denote the excited states by the quantum numbers of the excited electron, assuming the other one remains in the $1s$ shell. We do not include the continuum states, since we are interested in the ATAS features below the ionization potential. The Hamiltonian $H=H_0+V_L(t)$ consists of a field-free atomic part and the laser-matter interaction, with matrix elements given by
\begin{eqnarray}\langle i|H_0|j\rangle&=&E_i\delta_{ij}\\
\langle i|V_L(t)|j\rangle&=&-\langle i|\mathbf{d}|j\rangle\cdot\mathbf{E}(t)\end{eqnarray}
where $\mathbf{d}$ is the dipole operator, $\mathbf{E}(t)$ is the electric field and $E_i$ are the field-free energies of the states. The matrix elements are calculated in the quantum defect picture using field-free energies from the NIST database \cite{mayer2022,Yatsenko:1999aa,NIST_ASD}. To account for the decay of the excited states we include a diagonal term in the Hamiltonian $-\text{i}\Gamma/2$, where $\tau=2/\Gamma\simeq27$ fs. We expand the wavefunction over all magnetic quantum numbers $m\in[-\ell,\ell]$. The NIR and XUV fields are written in general form as
\begin{equation}\label{eq:ellflds}
    \mathbf{E}(t)=\frac{\mathcal{E}(t)}{\sqrt{2(1+\epsilon^2)}}\left[\frac{1+\epsilon}{2}\mathbf{e}_++\frac{1-\epsilon}{2}\mathbf{e}_-\right]e^{-\text{i}\omega t}+\text{c.c.}
\end{equation}
where $\epsilon$ is the ellipticity, $\mathbf{e}_\pm=(\hat{x}\mp\text{i}\hat{y})/\sqrt{2}$ the right-/left-circular basis, respectively, and $\epsilon=\pm1$ corresponds respectively to right/left-circular polarization.
We assume the initial wavefunction at the start of the simulation $t=t_0$ to be in the ground state $|\Psi(t_0)\rangle=|1s^2\rangle$ and obtain numerically the time-dependent amplitudes $c_i(t)$, from which the time-dependent dipole moment $\mathbf{d}(t)=\sum_{ij}c^*_i(t)c_j(t)\langle i|\mathbf{d}|j\rangle$ and the ATAS single-atom response $S(\omega)=2 \omega \text{Im}\left[\tilde{\mathbf{d}}(\omega)\cdot\tilde{\mathbf{E}}^*(\omega)\right]$ can be obtained. The change in signal $\Delta S$ is obtained by subtracting the pre-time zero signal $S_0$, where the NIR field can not modify the population, from all delays. The experimental observable change in absorbance for dilute gases is proportional to the change in single-atom response through scaling with the gas density $\rho$: $\Delta A \propto \frac{\rho}{\epsilon_0 c} \Delta S$, with the vacuum permittivity $\epsilon_0$ and the speed of light $c$~\cite{wu2016}.

\section*{Appendix C: Perturbation theory}

Here we derive the perturbative expression for the circular dichroism of the dipole moment of an atom in the presence of elliptically polarized XUV and NIR pulses, where we take the hydrogenic approximation for helium in a quantum defect picture \cite{mayer2022,Yatsenko:1999aa,NIST_ASD} as in the effective few-level model. The fields are given in the helical basis as in Eq. \ref{eq:ellflds}.
In first-order perturbation theory the absorption of an XUV photon from the ground state with $\ell=0$ angular momentum leads to a superposition of excited states
\begin{equation}|\Psi_{e}(t)\rangle=c_{p_{m=1}}(t)|p_{m=1}\rangle+c_{p_{m=-1}}(t)|p_{m=-1}\rangle\end{equation}
where for simplicity we focus on a given state of the $1snp$ series.
In the Rotating Wave Approximation (RWA) and the short pulse limit $\mathcal{E}_X(t)=F_X\delta(t)$,
\begin{eqnarray}c_{p_{m=\pm1}}(t)&=&\text{i}\frac{1\mp\epsilon_X}{\sqrt{2(1+\epsilon_X^2)}}\int_{-\infty}^{t}dt'\frac{d_{s,p_\pm}}{2}\mathcal{E}_X(t')e^{-\text{i}\Delta_{gp}t'}\nonumber\\
&&=\text{i}\Theta(t)\frac{1\mp\epsilon_X}{\sqrt{2(1+\epsilon_X^2)}}\frac{F_Xd_{s,p_\pm}}{2}\end{eqnarray}
where $\Theta(t)$ is the Heaviside function, $\epsilon_X$ is the ellipticity of the XUV pulse, $d_{s,p_\pm}=\langle \Psi_g|d|np_\pm\rangle$ is the dipole matrix element between ground state and the 1s$n$p state with $\ell=1,\,m=\pm1$ and $\Delta_{gp}=E_{np}-E_{g}-\omega_X$ is the detuning between the ground and 1s$n$p excited state Eigenenergies with $\omega_X$ the central frequency of the XUV pulse. It follows that for a perfectly circularly polarized XUV pulse ($\epsilon_X=\pm1$) only either $|p_{m=-1}\rangle$ or $|p_{m=1}\rangle$ is populated, respectively.

Assuming a weak NIR pulse and retaining only one photon off-resonant transitions, we model its effect via a modulation by additional phases (due to the AC Stark shift) of the amplitudes. We consider the $n$p series ($\ell=1$) coupled to the $n$s ($\ell=0$) and $n$d ($\ell=2$) ones via a one-photon transition in accordance with the selection rules $\Delta\ell=\pm1$, $\Delta m=0,\pm1$. We obtain the wavefunction:
\begin{eqnarray}|\Psi_e(t)\rangle&=&c_{p_{+1}}\left(e^{\text{i}\phi_{++}(t)+\text{i}\phi_{+-}(t)}\right)|p_{+1}\rangle+\nonumber\\
&&c_{p_{-1}}\left(e^{\text{i}\phi_{-+}(t)+\text{i}\phi_{--}(t)}\right)|p_{-1},\rangle\end{eqnarray}
where the phases $\phi_{XL}$ are sorted by the helicity ($+$/$-$) of the XUV ($X$) and NIR ($L$) pulses, respectively:
\begin{eqnarray}\phi_{++}(t)&=&\frac{(1-\epsilon_L)^2}{2(1+\epsilon_L^2)}\sum_n \left(\Delta\varphi_{p_{+1},nd_2}+\Delta\varphi_{p_{+1},ns}\right)\nonumber\\
\phi_{+-}(t)&=&\frac{(1+\epsilon_L)^2}{2(1+\epsilon_L^2)}\sum_n\left(\Delta\varphi_{p_{-1},nd_0}+\Delta\varphi_{p_{-1},ns}\right)\nonumber\\
\phi_{-+}(t)&=&\frac{(1-\epsilon_L)^2}{2(1+\epsilon_L^2)}\sum_n\left(\Delta\varphi_{p_{+1},nd_0}+\Delta\varphi_{p_{+1},ns}\right)\nonumber\\
\phi_{--}(t)&=&\frac{(1+\epsilon_L)^2}{2(1+\epsilon_L^2)}\sum_n\left(\Delta\varphi_{p_{-1},nd_{-2}}+\Delta\varphi_{p_{-1},ns}\right)
\end{eqnarray}
and the light-induced phase $\Delta \varphi$ due to the Stark shift depends explicitly on atomic parameters as \cite{Delone:1999aa,Chini:2012aa}:
\begin{equation}\Delta\varphi_{p_{m},n\ell_{m'}}=\frac{1}{2}|d_{p_m,n\ell_m'}|^2\left(\frac{\Delta_{pn}}{\Delta_{pn}^2-\omega_L^2}\right)\int_{0}^{t}dt'\mathcal{E}^2_L(t'),\end{equation}
where $\omega_L$ is the NIR frequency, $\mathcal{E}^2_L(t)$ the NIR intensity and $\Delta_{pn} = E_p-E_n$.
The dipole matrix elements are given by
\begin{eqnarray}d_{p_m,n\ell_j}&=&\langle p_m|rY_{1,m-j}|n\ell_j\rangle=\nonumber\\
&&\sqrt{\frac{9(2\ell+1)}{4\pi}}\left(\begin{matrix}1 & 1 & \ell \\ 0 & 0 & 0\end{matrix}\right)\left(\begin{matrix} 1 & 1 & \ell \\ -m & m-j & j\end{matrix}\right)\nonumber\\
&&\times\int_0^{\infty}dr R_{n1}(r)rR_{n\ell}(r),\end{eqnarray}
where $R_{n1}(r)$ and $R_{n\ell}(r)$ are the radial part of the hydrogenic wavefunctions $\Psi_{nlm}(\mathbf{r})=R_{nl}(r)Y_{lm}(\theta,\phi)$, where $R_{n1}(r)$ is the radial wavefunction of the $p$ state with $\ell=1$. Wigner's 3j-symbol for $m_1=m_2=m_3=0$ is non-zero only if $\ell+\ell'+\ell''$ is an even integer, and since
\begin{equation}\left(\begin{matrix} \ell & \ell' & \ell'' \\ m & m' & m''\end{matrix}\right)=(-1)^{\ell+\ell'+\ell''}\left(\begin{matrix}\ell & \ell' & \ell'' \\ -m & -m' & -m''\end{matrix}\right)\end{equation}
it follows that $S_{p_{+1},nd_2}=S_{p_{-1},nd_{-2}}$ and $S_{p_{+1},ns}=S_{p_{-1},ns}$. The system is therefore parity symmetric and is left unchanged by a flip of both XUV and NIR ellipticities, in contrast to a parity-breaking medium like e.g. an aligned ensemble of chiral molecules, where the flip of XUV and NIR ellipticities will lead to sizeable differences similar to the effects studied in Ref. \cite{drescher2020}.

The ATAS response is given by $S(\omega)=\mathrm{Im}\left[\tilde{\mathbf{d}}(\omega)\cdot\tilde{\mathbf{E}}_{XUV}^*(\omega)\right]$, where
\begin{equation}\tilde{\mathbf{E}}^*_{XUV}(\omega)=\frac{\tilde{\mathcal{E}}^*(\omega)}{\sqrt{2(1+\epsilon_X^2)}}\left(\frac{1+\epsilon_X}{2}\mathbf{e}_-+\frac{1-\epsilon_X}{2}\mathbf{e}_+\right)+c.c.\end{equation}
with $\tilde{\mathcal{E}}(\omega)=\int dt\mathcal{E}(t)\exp(\text{i}(\omega-\omega_X)t)$ and similar $\tilde{\mathbf{d}}(\omega)$ is obtained through Fourier transformation from the time-dependent dipole moment $\mathbf{d}(t)=\langle\Psi(t)|\mathbf{d}|\Psi(t)\rangle=\langle\Psi_g|\mathbf{d}|\Psi_e(t)\rangle+c.c.$.
The ATAS dichroism is obtained by substracting the signal with RCP NIR from the LCP NIR one, where the XUV ellipticity is fixed
\begin{equation}
    \text{CD}(\omega)=\mathrm{Im}\left[(\tilde{\mathbf{d}}_{RCP}(\omega)-\tilde{\mathbf{d}}_{LCP}(\omega))\cdot\tilde{\mathbf{E}}^*_{XUV}(\omega)\right]
    \end{equation}

The time-dependent dipole moment for each NIR helicity is given by summation over the RCP/LCP ($\alpha=\pm$) contributions:
\begin{eqnarray}\mathbf{d}_{RCP}(t)&=&\frac{\text{i}\Theta(t)F_X}{2\sqrt{2(1+\epsilon_X^2)}}\sum_{\alpha=\pm}|d_{s,p_{\alpha}}|^2\nonumber\\
&&\times\left(1+\alpha\epsilon_X\right)e^{\text{i}\phi_{\alpha -}(t)}\mathbf{e}_{\alpha}+c.c.\end{eqnarray}
\begin{eqnarray}\mathbf{d}_{LCP}(t)&=&\frac{\text{i}\Theta(t)F_X}{2\sqrt{2(1+\epsilon_X^2)}}\sum_{\alpha=\pm}|d_{s,p_{\alpha}}|^2\nonumber\\
&&\times\left(1+\alpha\epsilon_X\right)e^{\text{i}\phi_{\alpha +}(t)}\mathbf{e}_{\alpha}+c.c.\end{eqnarray}
and $\Delta\mathbf{d}(t)=\mathbf{d}_{RCP}(t)-\mathbf{d}_{LCP}(t)$ is given by
\begin{eqnarray}\Delta\mathbf{d}(t)&=&\frac{\text{i}\Theta(t)F_X}{2\sqrt{2(1+\epsilon_X^2)}}\nonumber\\
&&\sum_{\alpha=\pm}|d_{s,p_\alpha}|^2(1+\alpha\epsilon_X)\nonumber\\
&&\times\left(e^{\text{i}\phi_{\alpha -}(t)}-e^{\text{i}\phi_{\alpha +}(t)}\right)\mathbf{e}_{\alpha}+c.c.\end{eqnarray}
By transforming to frequency domain by Fourier transformation, denoting that $d_{s,p_{+1}}=d_{s,p_{-1}}=d_{s,p}$ and introducing $A_\alpha(\omega)=\int_0^\infty\left(e^{\text{i}\phi_{\alpha -}(t)}-e^{\text{i}\phi_{\alpha +}(t)}\right)e^{\text{i}\omega t}$, using $A_+(\omega)=-A_-(\omega)$, $\mathbf{e}_+\cdot\mathbf{e}_-=1$ and the definition of $\tilde{\mathbf{E}}^*_{XUV}(\omega)$ from above, the circular dichroism is given by
\begin{equation}\mathrm{CD}(\omega)=2|d_{s,p}|^2\frac{\epsilon_X}{1+\epsilon^2_X}\mathrm{Re}\left[A(\omega)\tilde{E}^*(\omega)\right].\end{equation}
We see that for a linearly polarized XUV pulse along the x-axis  $\epsilon_X=0$, the circular dichroism is zero because the dichroic dipole moment $\Delta\mathbf{d}(t)$ is oriented along the y-axis. For opposite XUV ellipticities, $\mathrm{CD}_{\epsilon_X}(\omega)=-\mathrm{CD}_{-\epsilon_X}(\omega)$, as expected from parity. For parity-broken systems $\mathrm{CD}_{\epsilon_X}(\omega)\neq \mathrm{CD}_{-\epsilon_X}(\omega)$.

\bibliography{main}

\begin{thebibliography}{39}%
\makeatletter
\providecommand \@ifxundefined [1]{%
 \@ifx{#1\undefined}
}%
\providecommand \@ifnum [1]{%
 \ifnum #1\expandafter \@firstoftwo
 \else \expandafter \@secondoftwo
 \fi
}%
\providecommand \@ifx [1]{%
 \ifx #1\expandafter \@firstoftwo
 \else \expandafter \@secondoftwo
 \fi
}%
\providecommand \natexlab [1]{#1}%
\providecommand \enquote  [1]{``#1''}%
\providecommand \bibnamefont  [1]{#1}%
\providecommand \bibfnamefont [1]{#1}%
\providecommand \citenamefont [1]{#1}%
\providecommand \href@noop [0]{\@secondoftwo}%
\providecommand \href [0]{\begingroup \@sanitize@url \@href}%
\providecommand \@href[1]{\@@startlink{#1}\@@href}%
\providecommand \@@href[1]{\endgroup#1\@@endlink}%
\providecommand \@sanitize@url [0]{\catcode `\\12\catcode `\$12\catcode
  `\&12\catcode `\#12\catcode `\^12\catcode `\_12\catcode `\%12\relax}%
\providecommand \@@startlink[1]{}%
\providecommand \@@endlink[0]{}%
\providecommand \url  [0]{\begingroup\@sanitize@url \@url }%
\providecommand \@url [1]{\endgroup\@href {#1}{\urlprefix }}%
\providecommand \urlprefix  [0]{URL }%
\providecommand \Eprint [0]{\href }%
\providecommand \doibase [0]{https://doi.org/}%
\providecommand \selectlanguage [0]{\@gobble}%
\providecommand \bibinfo  [0]{\@secondoftwo}%
\providecommand \bibfield  [0]{\@secondoftwo}%
\providecommand \translation [1]{[#1]}%
\providecommand \BibitemOpen [0]{}%
\providecommand \bibitemStop [0]{}%
\providecommand \bibitemNoStop [0]{.\EOS\space}%
\providecommand \EOS [0]{\spacefactor3000\relax}%
\providecommand \BibitemShut  [1]{\csname bibitem#1\endcsname}%
\let\auto@bib@innerbib\@empty
\bibitem [{\citenamefont {Leone}\ \emph {et~al.}(2014)\citenamefont {Leone},
  \citenamefont {McCurdy}, \citenamefont {Burgd{\"o}rfer}, \citenamefont
  {Cederbaum}, \citenamefont {Chang}, \citenamefont {Dudovich}, \citenamefont
  {Feist}, \citenamefont {Greene}, \citenamefont {Ivanov}, \citenamefont
  {Kienberger}, \citenamefont {Keller}, \citenamefont {Kling}, \citenamefont
  {Loh}, \citenamefont {Pfeifer}, \citenamefont {Pfeiffer}, \citenamefont
  {Santra}, \citenamefont {Schafer}, \citenamefont {Stolow}, \citenamefont
  {Thumm},\ and\ \citenamefont {Vrakking}}]{leone2014}%
  \BibitemOpen
  \bibfield  {author} {\bibinfo {author} {\bibfnamefont {S.~R.}\ \bibnamefont
  {Leone}}, \bibinfo {author} {\bibfnamefont {C.~W.}\ \bibnamefont {McCurdy}},
  \bibinfo {author} {\bibfnamefont {J.}~\bibnamefont {Burgd{\"o}rfer}},
  \bibinfo {author} {\bibfnamefont {L.~S.}\ \bibnamefont {Cederbaum}}, \bibinfo
  {author} {\bibfnamefont {Z.}~\bibnamefont {Chang}}, \bibinfo {author}
  {\bibfnamefont {N.}~\bibnamefont {Dudovich}}, \bibinfo {author}
  {\bibfnamefont {J.}~\bibnamefont {Feist}}, \bibinfo {author} {\bibfnamefont
  {C.~H.}\ \bibnamefont {Greene}}, \bibinfo {author} {\bibfnamefont
  {M.}~\bibnamefont {Ivanov}}, \bibinfo {author} {\bibfnamefont
  {R.}~\bibnamefont {Kienberger}}, \bibinfo {author} {\bibfnamefont
  {U.}~\bibnamefont {Keller}}, \bibinfo {author} {\bibfnamefont {M.~F.}\
  \bibnamefont {Kling}}, \bibinfo {author} {\bibfnamefont {Z.-H.}\ \bibnamefont
  {Loh}}, \bibinfo {author} {\bibfnamefont {T.}~\bibnamefont {Pfeifer}},
  \bibinfo {author} {\bibfnamefont {A.~N.}\ \bibnamefont {Pfeiffer}}, \bibinfo
  {author} {\bibfnamefont {R.}~\bibnamefont {Santra}}, \bibinfo {author}
  {\bibfnamefont {K.}~\bibnamefont {Schafer}}, \bibinfo {author} {\bibfnamefont
  {A.}~\bibnamefont {Stolow}}, \bibinfo {author} {\bibfnamefont
  {U.}~\bibnamefont {Thumm}},\ and\ \bibinfo {author} {\bibfnamefont
  {M.~J.~J.}\ \bibnamefont {Vrakking}},\ }\href
  {https://doi.org/10.1038/nphoton.2014.48} {\bibfield  {journal} {\bibinfo
  {journal} {Nature Photonics}\ }\textbf {\bibinfo {volume} {8}},\ \bibinfo
  {pages} {162} (\bibinfo {year} {2014})}\BibitemShut {NoStop}%
\bibitem [{\citenamefont {Willems}\ \emph {et~al.}(2015)\citenamefont
  {Willems}, \citenamefont {Smeenk}, \citenamefont {Zhavoronkov}, \citenamefont
  {Kornilov}, \citenamefont {Radu}, \citenamefont {Schmidbauer}, \citenamefont
  {Hanke}, \citenamefont {{von Korff Schmising}}, \citenamefont {Vrakking},\
  and\ \citenamefont {Eisebitt}}]{willems2015}%
  \BibitemOpen
  \bibfield  {author} {\bibinfo {author} {\bibfnamefont {F.}~\bibnamefont
  {Willems}}, \bibinfo {author} {\bibfnamefont {C.~T.~L.}\ \bibnamefont
  {Smeenk}}, \bibinfo {author} {\bibfnamefont {N.}~\bibnamefont {Zhavoronkov}},
  \bibinfo {author} {\bibfnamefont {O.}~\bibnamefont {Kornilov}}, \bibinfo
  {author} {\bibfnamefont {I.}~\bibnamefont {Radu}}, \bibinfo {author}
  {\bibfnamefont {M.}~\bibnamefont {Schmidbauer}}, \bibinfo {author}
  {\bibfnamefont {M.}~\bibnamefont {Hanke}}, \bibinfo {author} {\bibfnamefont
  {C.}~\bibnamefont {{von Korff Schmising}}}, \bibinfo {author} {\bibfnamefont
  {M.~J.~J.}\ \bibnamefont {Vrakking}},\ and\ \bibinfo {author} {\bibfnamefont
  {S.}~\bibnamefont {Eisebitt}},\ }\href
  {https://doi.org/10.1103/PhysRevB.92.220405} {\bibfield  {journal} {\bibinfo
  {journal} {Physical Review B}\ }\textbf {\bibinfo {volume} {92}},\ \bibinfo
  {pages} {220405} (\bibinfo {year} {2015})}\BibitemShut {NoStop}%
\bibitem [{\citenamefont {Siegrist}\ \emph {et~al.}(2019)\citenamefont
  {Siegrist}, \citenamefont {Gessner}, \citenamefont {Ossiander}, \citenamefont
  {Denker}, \citenamefont {Chang}, \citenamefont {Schr{\"o}der}, \citenamefont
  {Guggenmos}, \citenamefont {Cui}, \citenamefont {Walowski}, \citenamefont
  {Martens}, \citenamefont {Dewhurst}, \citenamefont {Kleineberg},
  \citenamefont {M{\"u}nzenberg}, \citenamefont {Sharma},\ and\ \citenamefont
  {Schultze}}]{siegrist2019}%
  \BibitemOpen
  \bibfield  {author} {\bibinfo {author} {\bibfnamefont {F.}~\bibnamefont
  {Siegrist}}, \bibinfo {author} {\bibfnamefont {J.~A.}\ \bibnamefont
  {Gessner}}, \bibinfo {author} {\bibfnamefont {M.}~\bibnamefont {Ossiander}},
  \bibinfo {author} {\bibfnamefont {C.}~\bibnamefont {Denker}}, \bibinfo
  {author} {\bibfnamefont {Y.-P.}\ \bibnamefont {Chang}}, \bibinfo {author}
  {\bibfnamefont {M.~C.}\ \bibnamefont {Schr{\"o}der}}, \bibinfo {author}
  {\bibfnamefont {A.}~\bibnamefont {Guggenmos}}, \bibinfo {author}
  {\bibfnamefont {Y.}~\bibnamefont {Cui}}, \bibinfo {author} {\bibfnamefont
  {J.}~\bibnamefont {Walowski}}, \bibinfo {author} {\bibfnamefont
  {U.}~\bibnamefont {Martens}}, \bibinfo {author} {\bibfnamefont {J.~K.}\
  \bibnamefont {Dewhurst}}, \bibinfo {author} {\bibfnamefont {U.}~\bibnamefont
  {Kleineberg}}, \bibinfo {author} {\bibfnamefont {M.}~\bibnamefont
  {M{\"u}nzenberg}}, \bibinfo {author} {\bibfnamefont {S.}~\bibnamefont
  {Sharma}},\ and\ \bibinfo {author} {\bibfnamefont {M.}~\bibnamefont
  {Schultze}},\ }\href {https://doi.org/10.1038/s41586-019-1333-x} {\bibfield
  {journal} {\bibinfo  {journal} {Nature}\ }\textbf {\bibinfo {volume} {571}},\
  \bibinfo {pages} {240} (\bibinfo {year} {2019})}\BibitemShut {NoStop}%
\bibitem [{\citenamefont {G{\'e}neaux}\ \emph {et~al.}(2024)\citenamefont
  {G{\'e}neaux}, \citenamefont {Chang}, \citenamefont {Guggenmos},
  \citenamefont {Delaunay}, \citenamefont {L{\'e}gar{\'e}}, \citenamefont
  {L{\'e}gar{\'e}}, \citenamefont {L{\"u}ning}, \citenamefont {Parpiiev},
  \citenamefont {Molesky}, \citenamefont {{de Roulet}}, \citenamefont {Zuerch},
  \citenamefont {Sharma}, \citenamefont {Schultze},\ and\ \citenamefont
  {Leone}}]{geneaux2024}%
  \BibitemOpen
  \bibfield  {author} {\bibinfo {author} {\bibfnamefont {R.}~\bibnamefont
  {G{\'e}neaux}}, \bibinfo {author} {\bibfnamefont {H.-T.}\ \bibnamefont
  {Chang}}, \bibinfo {author} {\bibfnamefont {A.}~\bibnamefont {Guggenmos}},
  \bibinfo {author} {\bibfnamefont {R.}~\bibnamefont {Delaunay}}, \bibinfo
  {author} {\bibfnamefont {F.}~\bibnamefont {L{\'e}gar{\'e}}}, \bibinfo
  {author} {\bibfnamefont {K.}~\bibnamefont {L{\'e}gar{\'e}}}, \bibinfo
  {author} {\bibfnamefont {J.}~\bibnamefont {L{\"u}ning}}, \bibinfo {author}
  {\bibfnamefont {T.}~\bibnamefont {Parpiiev}}, \bibinfo {author}
  {\bibfnamefont {I.~J.~P.}\ \bibnamefont {Molesky}}, \bibinfo {author}
  {\bibfnamefont {B.~R.}\ \bibnamefont {{de Roulet}}}, \bibinfo {author}
  {\bibfnamefont {M.~W.}\ \bibnamefont {Zuerch}}, \bibinfo {author}
  {\bibfnamefont {S.}~\bibnamefont {Sharma}}, \bibinfo {author} {\bibfnamefont
  {M.}~\bibnamefont {Schultze}},\ and\ \bibinfo {author} {\bibfnamefont
  {S.~R.}\ \bibnamefont {Leone}},\ }\href
  {https://doi.org/10.1103/PhysRevLett.133.106902} {\bibfield  {journal}
  {\bibinfo  {journal} {Physical Review Letters}\ }\textbf {\bibinfo {volume}
  {133}},\ \bibinfo {pages} {106902} (\bibinfo {year} {2024})}\BibitemShut
  {NoStop}%
\bibitem [{\citenamefont {Hasan}\ and\ \citenamefont {Kane}(2010)}]{hasan2010}%
  \BibitemOpen
  \bibfield  {author} {\bibinfo {author} {\bibfnamefont {M.~Z.}\ \bibnamefont
  {Hasan}}\ and\ \bibinfo {author} {\bibfnamefont {C.~L.}\ \bibnamefont
  {Kane}},\ }\href {https://doi.org/10.1103/RevModPhys.82.3045} {\bibfield
  {journal} {\bibinfo  {journal} {Reviews of Modern Physics}\ }\textbf
  {\bibinfo {volume} {82}},\ \bibinfo {pages} {3045} (\bibinfo {year}
  {2010})}\BibitemShut {NoStop}%
\bibitem [{\citenamefont {Schaibley}\ \emph {et~al.}(2016)\citenamefont
  {Schaibley}, \citenamefont {Yu}, \citenamefont {Clark}, \citenamefont
  {Rivera}, \citenamefont {Ross}, \citenamefont {Seyler}, \citenamefont {Yao},\
  and\ \citenamefont {Xu}}]{schaibley2016}%
  \BibitemOpen
  \bibfield  {author} {\bibinfo {author} {\bibfnamefont {J.~R.}\ \bibnamefont
  {Schaibley}}, \bibinfo {author} {\bibfnamefont {H.}~\bibnamefont {Yu}},
  \bibinfo {author} {\bibfnamefont {G.}~\bibnamefont {Clark}}, \bibinfo
  {author} {\bibfnamefont {P.}~\bibnamefont {Rivera}}, \bibinfo {author}
  {\bibfnamefont {J.~S.}\ \bibnamefont {Ross}}, \bibinfo {author}
  {\bibfnamefont {K.~L.}\ \bibnamefont {Seyler}}, \bibinfo {author}
  {\bibfnamefont {W.}~\bibnamefont {Yao}},\ and\ \bibinfo {author}
  {\bibfnamefont {X.}~\bibnamefont {Xu}},\ }\href
  {https://doi.org/10.1038/natrevmats.2016.55} {\bibfield  {journal} {\bibinfo
  {journal} {Nature Reviews Materials}\ }\textbf {\bibinfo {volume} {1}},\
  \bibinfo {pages} {1} (\bibinfo {year} {2016})}\BibitemShut {NoStop}%
\bibitem [{\citenamefont {Bao}\ \emph {et~al.}(2021)\citenamefont {Bao},
  \citenamefont {Tang}, \citenamefont {Sun},\ and\ \citenamefont
  {Zhou}}]{bao2021}%
  \BibitemOpen
  \bibfield  {author} {\bibinfo {author} {\bibfnamefont {C.}~\bibnamefont
  {Bao}}, \bibinfo {author} {\bibfnamefont {P.}~\bibnamefont {Tang}}, \bibinfo
  {author} {\bibfnamefont {D.}~\bibnamefont {Sun}},\ and\ \bibinfo {author}
  {\bibfnamefont {S.}~\bibnamefont {Zhou}},\ }\href
  {https://doi.org/10.1038/s42254-021-00388-1} {\bibfield  {journal} {\bibinfo
  {journal} {Nature Reviews Physics}\ ,\ \bibinfo {pages} {1}} (\bibinfo {year}
  {2021})}\BibitemShut {NoStop}%
\bibitem [{\citenamefont {Yang}\ \emph {et~al.}(2021)\citenamefont {Yang},
  \citenamefont {Naaman}, \citenamefont {Paltiel},\ and\ \citenamefont
  {Parkin}}]{yang2021}%
  \BibitemOpen
  \bibfield  {author} {\bibinfo {author} {\bibfnamefont {S.-H.}\ \bibnamefont
  {Yang}}, \bibinfo {author} {\bibfnamefont {R.}~\bibnamefont {Naaman}},
  \bibinfo {author} {\bibfnamefont {Y.}~\bibnamefont {Paltiel}},\ and\ \bibinfo
  {author} {\bibfnamefont {S.~S.~P.}\ \bibnamefont {Parkin}},\ }\href
  {https://doi.org/10.1038/s42254-021-00302-9} {\bibfield  {journal} {\bibinfo
  {journal} {Nature Reviews Physics}\ }\textbf {\bibinfo {volume} {3}},\
  \bibinfo {pages} {328} (\bibinfo {year} {2021})}\BibitemShut {NoStop}%
\bibitem [{\citenamefont {Rouxel}\ and\ \citenamefont
  {Mukamel}(2022)}]{rouxel2022}%
  \BibitemOpen
  \bibfield  {author} {\bibinfo {author} {\bibfnamefont {J.~R.}\ \bibnamefont
  {Rouxel}}\ and\ \bibinfo {author} {\bibfnamefont {S.}~\bibnamefont
  {Mukamel}},\ }\href {https://doi.org/10.1021/acs.chemrev.2c00115} {\bibfield
  {journal} {\bibinfo  {journal} {Chemical Reviews}\ }\textbf {\bibinfo
  {volume} {122}},\ \bibinfo {pages} {16802} (\bibinfo {year}
  {2022})}\BibitemShut {NoStop}%
\bibitem [{\citenamefont {Franzen}\ and\ \citenamefont
  {Emslie}(1957)}]{franzen1957}%
  \BibitemOpen
  \bibfield  {author} {\bibinfo {author} {\bibfnamefont {W.}~\bibnamefont
  {Franzen}}\ and\ \bibinfo {author} {\bibfnamefont {A.~G.}\ \bibnamefont
  {Emslie}},\ }\href {https://doi.org/10.1103/PhysRev.108.1453} {\bibfield
  {journal} {\bibinfo  {journal} {Physical Review}\ }\textbf {\bibinfo {volume}
  {108}},\ \bibinfo {pages} {1453} (\bibinfo {year} {1957})}\BibitemShut
  {NoStop}%
\bibitem [{\citenamefont {Skrotskii}\ and\ \citenamefont
  {Izyumova}(1961)}]{skrotskii1961}%
  \BibitemOpen
  \bibfield  {author} {\bibinfo {author} {\bibfnamefont {G.~V.}\ \bibnamefont
  {Skrotskii}}\ and\ \bibinfo {author} {\bibfnamefont {T.~G.}\ \bibnamefont
  {Izyumova}},\ }\href {https://doi.org/10.1070/PU1961v004n02ABEH003331}
  {\bibfield  {journal} {\bibinfo  {journal} {Soviet Physics Uspekhi}\ }\textbf
  {\bibinfo {volume} {4}},\ \bibinfo {pages} {177} (\bibinfo {year}
  {1961})}\BibitemShut {NoStop}%
\bibitem [{\citenamefont {Meier}\ and\ \citenamefont
  {Zakharchenya}(1984)}]{meier1984}%
  \BibitemOpen
  \bibfield  {author} {\bibinfo {author} {\bibfnamefont {F.}~\bibnamefont
  {Meier}}\ and\ \bibinfo {author} {\bibfnamefont {{\relax BP}.}~\bibnamefont
  {Zakharchenya}},\ }\href@noop {} {\emph {\bibinfo {title} {Optical
  Orientation}}},\ Modern {{Problems}} in {{Condensed Matter Sciences}}\
  (\bibinfo  {publisher} {Elsevier},\ \bibinfo {address} {Burlington, MA},\
  \bibinfo {year} {1984})\BibitemShut {NoStop}%
\bibitem [{\citenamefont {Sutcliffe}\ \emph {et~al.}(2024)\citenamefont
  {Sutcliffe}, \citenamefont {Kazmierczak},\ and\ \citenamefont
  {Hadt}}]{sutcliffe2024}%
  \BibitemOpen
  \bibfield  {author} {\bibinfo {author} {\bibfnamefont {E.}~\bibnamefont
  {Sutcliffe}}, \bibinfo {author} {\bibfnamefont {N.~P.}\ \bibnamefont
  {Kazmierczak}},\ and\ \bibinfo {author} {\bibfnamefont {R.~G.}\ \bibnamefont
  {Hadt}},\ }\href {https://doi.org/10.1126/science.ads0512} {\bibfield
  {journal} {\bibinfo  {journal} {Science}\ }\textbf {\bibinfo {volume}
  {386}},\ \bibinfo {pages} {888} (\bibinfo {year} {2024})}\BibitemShut
  {NoStop}%
\bibitem [{\citenamefont {Wu}\ \emph {et~al.}(2016)\citenamefont {Wu},
  \citenamefont {Chen}, \citenamefont {Camp}, \citenamefont {Schafer},\ and\
  \citenamefont {Gaarde}}]{wu2016}%
  \BibitemOpen
  \bibfield  {author} {\bibinfo {author} {\bibfnamefont {M.}~\bibnamefont
  {Wu}}, \bibinfo {author} {\bibfnamefont {S.}~\bibnamefont {Chen}}, \bibinfo
  {author} {\bibfnamefont {S.}~\bibnamefont {Camp}}, \bibinfo {author}
  {\bibfnamefont {K.~J.}\ \bibnamefont {Schafer}},\ and\ \bibinfo {author}
  {\bibfnamefont {M.~B.}\ \bibnamefont {Gaarde}},\ }\href
  {https://doi.org/10.1088/0953-4075/49/6/062003} {\bibfield  {journal}
  {\bibinfo  {journal} {Journal of Physics B: Atomic, Molecular and Optical
  Physics}\ }\textbf {\bibinfo {volume} {49}},\ \bibinfo {pages} {062003}
  (\bibinfo {year} {2016})}\BibitemShut {NoStop}%
\bibitem [{\citenamefont {Chini}\ \emph {et~al.}(2012)\citenamefont {Chini},
  \citenamefont {Zhao}, \citenamefont {Wang}, \citenamefont {Cheng},
  \citenamefont {Hu},\ and\ \citenamefont {Chang}}]{Chini:2012aa}%
  \BibitemOpen
  \bibfield  {author} {\bibinfo {author} {\bibfnamefont {M.}~\bibnamefont
  {Chini}}, \bibinfo {author} {\bibfnamefont {B.}~\bibnamefont {Zhao}},
  \bibinfo {author} {\bibfnamefont {H.}~\bibnamefont {Wang}}, \bibinfo {author}
  {\bibfnamefont {Y.}~\bibnamefont {Cheng}}, \bibinfo {author} {\bibfnamefont
  {S.}~\bibnamefont {Hu}},\ and\ \bibinfo {author} {\bibfnamefont
  {Z.}~\bibnamefont {Chang}},\ }\href
  {https://doi.org/https://doi.org/10.1103/PhysRevLett.109.073601} {\bibfield
  {journal} {\bibinfo  {journal} {Phys. Rev. Lett.}\ }\textbf {\bibinfo
  {volume} {109}},\ \bibinfo {pages} {073601} (\bibinfo {year}
  {2012})}\BibitemShut {NoStop}%
\bibitem [{\citenamefont {Reduzzi}\ \emph {et~al.}(2015)\citenamefont
  {Reduzzi}, \citenamefont {Hummert}, \citenamefont {Dubrouil}, \citenamefont
  {Calegari}, \citenamefont {Nisoli}, \citenamefont {Frassetto}, \citenamefont
  {Poletto}, \citenamefont {Chen}, \citenamefont {Wu}, \citenamefont {Gaarde},
  \citenamefont {Schafer},\ and\ \citenamefont {Sansone}}]{reduzzi2015}%
  \BibitemOpen
  \bibfield  {author} {\bibinfo {author} {\bibfnamefont {M.}~\bibnamefont
  {Reduzzi}}, \bibinfo {author} {\bibfnamefont {J.}~\bibnamefont {Hummert}},
  \bibinfo {author} {\bibfnamefont {A.}~\bibnamefont {Dubrouil}}, \bibinfo
  {author} {\bibfnamefont {F.}~\bibnamefont {Calegari}}, \bibinfo {author}
  {\bibfnamefont {M.}~\bibnamefont {Nisoli}}, \bibinfo {author} {\bibfnamefont
  {F.}~\bibnamefont {Frassetto}}, \bibinfo {author} {\bibfnamefont
  {L.}~\bibnamefont {Poletto}}, \bibinfo {author} {\bibfnamefont
  {S.}~\bibnamefont {Chen}}, \bibinfo {author} {\bibfnamefont {M.}~\bibnamefont
  {Wu}}, \bibinfo {author} {\bibfnamefont {M.~B.}\ \bibnamefont {Gaarde}},
  \bibinfo {author} {\bibfnamefont {K.}~\bibnamefont {Schafer}},\ and\ \bibinfo
  {author} {\bibfnamefont {G.}~\bibnamefont {Sansone}},\ }\href
  {https://doi.org/10.1103/PhysRevA.92.033408} {\bibfield  {journal} {\bibinfo
  {journal} {Physical Review A}\ }\textbf {\bibinfo {volume} {92}},\ \bibinfo
  {pages} {033408} (\bibinfo {year} {2015})}\BibitemShut {NoStop}%
\bibitem [{\citenamefont {Hartung}\ \emph {et~al.}(2016)\citenamefont
  {Hartung}, \citenamefont {Morales}, \citenamefont {Kunitski}, \citenamefont
  {Henrichs}, \citenamefont {Laucke}, \citenamefont {Richter}, \citenamefont
  {Jahnke}, \citenamefont {Kalinin}, \citenamefont {Sch{\"o}ffler},
  \citenamefont {Schmidt}, \citenamefont {Ivanov}, \citenamefont {Smirnova},\
  and\ \citenamefont {D{\"o}rner}}]{hartung2016}%
  \BibitemOpen
  \bibfield  {author} {\bibinfo {author} {\bibfnamefont {A.}~\bibnamefont
  {Hartung}}, \bibinfo {author} {\bibfnamefont {F.}~\bibnamefont {Morales}},
  \bibinfo {author} {\bibfnamefont {M.}~\bibnamefont {Kunitski}}, \bibinfo
  {author} {\bibfnamefont {K.}~\bibnamefont {Henrichs}}, \bibinfo {author}
  {\bibfnamefont {A.}~\bibnamefont {Laucke}}, \bibinfo {author} {\bibfnamefont
  {M.}~\bibnamefont {Richter}}, \bibinfo {author} {\bibfnamefont
  {T.}~\bibnamefont {Jahnke}}, \bibinfo {author} {\bibfnamefont
  {A.}~\bibnamefont {Kalinin}}, \bibinfo {author} {\bibfnamefont
  {M.}~\bibnamefont {Sch{\"o}ffler}}, \bibinfo {author} {\bibfnamefont
  {L.~P.~H.}\ \bibnamefont {Schmidt}}, \bibinfo {author} {\bibfnamefont
  {M.}~\bibnamefont {Ivanov}}, \bibinfo {author} {\bibfnamefont
  {O.}~\bibnamefont {Smirnova}},\ and\ \bibinfo {author} {\bibfnamefont
  {R.}~\bibnamefont {D{\"o}rner}},\ }\href
  {https://doi.org/10.1038/nphoton.2016.109} {\bibfield  {journal} {\bibinfo
  {journal} {Nature Photonics}\ }\textbf {\bibinfo {volume} {10}},\ \bibinfo
  {pages} {526} (\bibinfo {year} {2016})}\BibitemShut {NoStop}%
\bibitem [{\citenamefont {{Jim{\'e}nez-Gal{\'a}n}}\ \emph
  {et~al.}(2018)\citenamefont {{Jim{\'e}nez-Gal{\'a}n}}, \citenamefont
  {Zhavoronkov}, \citenamefont {Ayuso}, \citenamefont {Morales}, \citenamefont
  {Patchkovskii}, \citenamefont {Schloz}, \citenamefont {Pisanty},
  \citenamefont {Smirnova},\ and\ \citenamefont {Ivanov}}]{jimenez-galan2018}%
  \BibitemOpen
  \bibfield  {author} {\bibinfo {author} {\bibfnamefont {{\'A}.}~\bibnamefont
  {{Jim{\'e}nez-Gal{\'a}n}}}, \bibinfo {author} {\bibfnamefont
  {N.}~\bibnamefont {Zhavoronkov}}, \bibinfo {author} {\bibfnamefont
  {D.}~\bibnamefont {Ayuso}}, \bibinfo {author} {\bibfnamefont
  {F.}~\bibnamefont {Morales}}, \bibinfo {author} {\bibfnamefont
  {S.}~\bibnamefont {Patchkovskii}}, \bibinfo {author} {\bibfnamefont
  {M.}~\bibnamefont {Schloz}}, \bibinfo {author} {\bibfnamefont
  {E.}~\bibnamefont {Pisanty}}, \bibinfo {author} {\bibfnamefont
  {O.}~\bibnamefont {Smirnova}},\ and\ \bibinfo {author} {\bibfnamefont
  {M.}~\bibnamefont {Ivanov}},\ }\href
  {https://doi.org/10.1103/PhysRevA.97.023409} {\bibfield  {journal} {\bibinfo
  {journal} {Physical Review A}\ }\textbf {\bibinfo {volume} {97}},\ \bibinfo
  {pages} {023409} (\bibinfo {year} {2018})}\BibitemShut {NoStop}%
\bibitem [{\citenamefont {Mayer}\ \emph {et~al.}(2022)\citenamefont {Mayer},
  \citenamefont {Beaulieu}, \citenamefont {{Jim{\'e}nez-Gal{\'a}n}},
  \citenamefont {Patchkovskii}, \citenamefont {Kornilov}, \citenamefont
  {Descamps}, \citenamefont {Petit}, \citenamefont {Smirnova}, \citenamefont
  {Mairesse},\ and\ \citenamefont {Ivanov}}]{mayer2022}%
  \BibitemOpen
  \bibfield  {author} {\bibinfo {author} {\bibfnamefont {N.}~\bibnamefont
  {Mayer}}, \bibinfo {author} {\bibfnamefont {S.}~\bibnamefont {Beaulieu}},
  \bibinfo {author} {\bibfnamefont {{\'A}.}~\bibnamefont
  {{Jim{\'e}nez-Gal{\'a}n}}}, \bibinfo {author} {\bibfnamefont
  {S.}~\bibnamefont {Patchkovskii}}, \bibinfo {author} {\bibfnamefont
  {O.}~\bibnamefont {Kornilov}}, \bibinfo {author} {\bibfnamefont
  {D.}~\bibnamefont {Descamps}}, \bibinfo {author} {\bibfnamefont
  {S.}~\bibnamefont {Petit}}, \bibinfo {author} {\bibfnamefont
  {O.}~\bibnamefont {Smirnova}}, \bibinfo {author} {\bibfnamefont
  {Y.}~\bibnamefont {Mairesse}},\ and\ \bibinfo {author} {\bibfnamefont
  {M.~Y.}\ \bibnamefont {Ivanov}},\ }\href
  {https://doi.org/10.1103/PhysRevLett.129.173202} {\bibfield  {journal}
  {\bibinfo  {journal} {Physical Review Letters}\ }\textbf {\bibinfo {volume}
  {129}},\ \bibinfo {pages} {173202} (\bibinfo {year} {2022})}\BibitemShut
  {NoStop}%
\bibitem [{\citenamefont {Carlstr{\"o}m}\ \emph {et~al.}(2024)\citenamefont
  {Carlstr{\"o}m}, \citenamefont {Tahouri}, \citenamefont {Papoulia},
  \citenamefont {Dahlstr{\"o}m}, \citenamefont {Ivanov}, \citenamefont
  {Smirnova},\ and\ \citenamefont {Patchkovskii}}]{carlstrom2024}%
  \BibitemOpen
  \bibfield  {author} {\bibinfo {author} {\bibfnamefont {S.}~\bibnamefont
  {Carlstr{\"o}m}}, \bibinfo {author} {\bibfnamefont {R.}~\bibnamefont
  {Tahouri}}, \bibinfo {author} {\bibfnamefont {A.}~\bibnamefont {Papoulia}},
  \bibinfo {author} {\bibfnamefont {J.~M.}\ \bibnamefont {Dahlstr{\"o}m}},
  \bibinfo {author} {\bibfnamefont {M.~Y.}\ \bibnamefont {Ivanov}}, \bibinfo
  {author} {\bibfnamefont {O.}~\bibnamefont {Smirnova}},\ and\ \bibinfo
  {author} {\bibfnamefont {S.}~\bibnamefont {Patchkovskii}},\ }\href
  {https://doi.org/10.48550/arXiv.2306.15665} {\bibinfo {title}
  {Spin-{{Polarized Photoelectrons}} in the {{Vicinity}} of {{Spectral
  Features}}}} (\bibinfo {year} {2024}),\ \Eprint
  {https://arxiv.org/abs/2306.15665} {arXiv:2306.15665 [physics]} \BibitemShut
  {NoStop}%
\bibitem [{\citenamefont {Eckart}\ \emph {et~al.}(2018)\citenamefont {Eckart},
  \citenamefont {Kunitski}, \citenamefont {Richter}, \citenamefont {Hartung},
  \citenamefont {Rist}, \citenamefont {Trinter}, \citenamefont {Fehre},
  \citenamefont {Schlott}, \citenamefont {Henrichs}, \citenamefont {Schmidt},
  \citenamefont {Jahnke}, \citenamefont {Sch{\"o}ffler}, \citenamefont {Liu},
  \citenamefont {Barth}, \citenamefont {Kaushal}, \citenamefont {Morales},
  \citenamefont {Ivanov}, \citenamefont {Smirnova},\ and\ \citenamefont
  {D{\"o}rner}}]{eckart2018}%
  \BibitemOpen
  \bibfield  {author} {\bibinfo {author} {\bibfnamefont {S.}~\bibnamefont
  {Eckart}}, \bibinfo {author} {\bibfnamefont {M.}~\bibnamefont {Kunitski}},
  \bibinfo {author} {\bibfnamefont {M.}~\bibnamefont {Richter}}, \bibinfo
  {author} {\bibfnamefont {A.}~\bibnamefont {Hartung}}, \bibinfo {author}
  {\bibfnamefont {J.}~\bibnamefont {Rist}}, \bibinfo {author} {\bibfnamefont
  {F.}~\bibnamefont {Trinter}}, \bibinfo {author} {\bibfnamefont
  {K.}~\bibnamefont {Fehre}}, \bibinfo {author} {\bibfnamefont
  {N.}~\bibnamefont {Schlott}}, \bibinfo {author} {\bibfnamefont
  {K.}~\bibnamefont {Henrichs}}, \bibinfo {author} {\bibfnamefont {L.~P.~H.}\
  \bibnamefont {Schmidt}}, \bibinfo {author} {\bibfnamefont {T.}~\bibnamefont
  {Jahnke}}, \bibinfo {author} {\bibfnamefont {M.}~\bibnamefont
  {Sch{\"o}ffler}}, \bibinfo {author} {\bibfnamefont {K.}~\bibnamefont {Liu}},
  \bibinfo {author} {\bibfnamefont {I.}~\bibnamefont {Barth}}, \bibinfo
  {author} {\bibfnamefont {J.}~\bibnamefont {Kaushal}}, \bibinfo {author}
  {\bibfnamefont {F.}~\bibnamefont {Morales}}, \bibinfo {author} {\bibfnamefont
  {M.}~\bibnamefont {Ivanov}}, \bibinfo {author} {\bibfnamefont
  {O.}~\bibnamefont {Smirnova}},\ and\ \bibinfo {author} {\bibfnamefont
  {R.}~\bibnamefont {D{\"o}rner}},\ }\href
  {https://doi.org/10.1038/s41567-018-0080-5} {\bibfield  {journal} {\bibinfo
  {journal} {Nature Physics}\ }\textbf {\bibinfo {volume} {14}},\ \bibinfo
  {pages} {701} (\bibinfo {year} {2018})}\BibitemShut {NoStop}%
\bibitem [{\citenamefont {De~Silva}\ \emph {et~al.}(2021)\citenamefont
  {De~Silva}, \citenamefont {Moon}, \citenamefont {Romans}, \citenamefont
  {Acharya}, \citenamefont {Dubey}, \citenamefont {Foster}, \citenamefont
  {Russ}, \citenamefont {Rischbieter}, \citenamefont {Douguet}, \citenamefont
  {Bartschat},\ and\ \citenamefont {Fischer}}]{desilva2021}%
  \BibitemOpen
  \bibfield  {author} {\bibinfo {author} {\bibfnamefont {A.~H. N.~C.}\
  \bibnamefont {De~Silva}}, \bibinfo {author} {\bibfnamefont {T.}~\bibnamefont
  {Moon}}, \bibinfo {author} {\bibfnamefont {K.~L.}\ \bibnamefont {Romans}},
  \bibinfo {author} {\bibfnamefont {B.~P.}\ \bibnamefont {Acharya}}, \bibinfo
  {author} {\bibfnamefont {S.}~\bibnamefont {Dubey}}, \bibinfo {author}
  {\bibfnamefont {K.}~\bibnamefont {Foster}}, \bibinfo {author} {\bibfnamefont
  {O.}~\bibnamefont {Russ}}, \bibinfo {author} {\bibfnamefont {C.}~\bibnamefont
  {Rischbieter}}, \bibinfo {author} {\bibfnamefont {N.}~\bibnamefont
  {Douguet}}, \bibinfo {author} {\bibfnamefont {K.}~\bibnamefont {Bartschat}},\
  and\ \bibinfo {author} {\bibfnamefont {D.}~\bibnamefont {Fischer}},\ }\href
  {https://doi.org/10.1103/PhysRevA.103.053125} {\bibfield  {journal} {\bibinfo
   {journal} {Physical Review A}\ }\textbf {\bibinfo {volume} {103}},\ \bibinfo
  {pages} {053125} (\bibinfo {year} {2021})}\BibitemShut {NoStop}%
\bibitem [{\citenamefont {Han}\ \emph {et~al.}(2023)\citenamefont {Han},
  \citenamefont {Ji}, \citenamefont {Bal{\v c}i{\=u}nas}, \citenamefont
  {Ueda},\ and\ \citenamefont {W{\"o}rner}}]{han2023}%
  \BibitemOpen
  \bibfield  {author} {\bibinfo {author} {\bibfnamefont {M.}~\bibnamefont
  {Han}}, \bibinfo {author} {\bibfnamefont {J.-B.}\ \bibnamefont {Ji}},
  \bibinfo {author} {\bibfnamefont {T.}~\bibnamefont {Bal{\v c}i{\=u}nas}},
  \bibinfo {author} {\bibfnamefont {K.}~\bibnamefont {Ueda}},\ and\ \bibinfo
  {author} {\bibfnamefont {H.~J.}\ \bibnamefont {W{\"o}rner}},\ }\href
  {https://doi.org/10.1038/s41567-022-01832-4} {\bibfield  {journal} {\bibinfo
  {journal} {Nature Physics}\ }\textbf {\bibinfo {volume} {19}},\ \bibinfo
  {pages} {230} (\bibinfo {year} {2023})}\BibitemShut {NoStop}%
\bibitem [{\citenamefont {Kheifets}(2024)}]{kheifets2024}%
  \BibitemOpen
  \bibfield  {author} {\bibinfo {author} {\bibfnamefont {A.~S.}\ \bibnamefont
  {Kheifets}},\ }\href {https://doi.org/10.1103/PhysRevResearch.6.L012002}
  {\bibfield  {journal} {\bibinfo  {journal} {Physical Review Research}\
  }\textbf {\bibinfo {volume} {6}},\ \bibinfo {pages} {L012002} (\bibinfo
  {year} {2024})}\BibitemShut {NoStop}%
\bibitem [{\citenamefont {Fano}(1985)}]{Fano:1985aa}%
  \BibitemOpen
  \bibfield  {author} {\bibinfo {author} {\bibfnamefont {U.}~\bibnamefont
  {Fano}},\ }\href {https://doi.org/https://doi.org/10.1103/PhysRevA.32.617}
  {\bibfield  {journal} {\bibinfo  {journal} {Physical Review A}\ }\textbf
  {\bibinfo {volume} {32}},\ \bibinfo {pages} {617} (\bibinfo {year}
  {1985})}\BibitemShut {NoStop}%
\bibitem [{\citenamefont {Bethe}\ and\ \citenamefont
  {Jackiw}(1997)}]{Bethe:1997aa}%
  \BibitemOpen
  \bibfield  {author} {\bibinfo {author} {\bibfnamefont {H.~A.}\ \bibnamefont
  {Bethe}}\ and\ \bibinfo {author} {\bibfnamefont {R.~W.}\ \bibnamefont
  {Jackiw}},\ }\href@noop {} {\emph {\bibinfo {title} {Intermediate quantum
  mechanics, Third Edition}}}\ (\bibinfo  {publisher} {CRC Press},\ \bibinfo
  {year} {1997})\ Chap.\ \bibinfo {chapter} {Chapter 11}\BibitemShut {NoStop}%
\bibitem [{\citenamefont {Busto}\ \emph {et~al.}(2019)\citenamefont {Busto},
  \citenamefont {Vinbladh}, \citenamefont {Zhong}, \citenamefont {Isinger},
  \citenamefont {Nandi}, \citenamefont {Maclot}, \citenamefont {Johnsson},
  \citenamefont {Gisselbrecht}, \citenamefont {L'Huillier}, \citenamefont
  {Lindroth},\ and\ \citenamefont {Dahlstr{\"o}m}}]{busto2019}%
  \BibitemOpen
  \bibfield  {author} {\bibinfo {author} {\bibfnamefont {D.}~\bibnamefont
  {Busto}}, \bibinfo {author} {\bibfnamefont {J.}~\bibnamefont {Vinbladh}},
  \bibinfo {author} {\bibfnamefont {S.}~\bibnamefont {Zhong}}, \bibinfo
  {author} {\bibfnamefont {M.}~\bibnamefont {Isinger}}, \bibinfo {author}
  {\bibfnamefont {S.}~\bibnamefont {Nandi}}, \bibinfo {author} {\bibfnamefont
  {S.}~\bibnamefont {Maclot}}, \bibinfo {author} {\bibfnamefont
  {P.}~\bibnamefont {Johnsson}}, \bibinfo {author} {\bibfnamefont
  {M.}~\bibnamefont {Gisselbrecht}}, \bibinfo {author} {\bibfnamefont
  {A.}~\bibnamefont {L'Huillier}}, \bibinfo {author} {\bibfnamefont
  {E.}~\bibnamefont {Lindroth}},\ and\ \bibinfo {author} {\bibfnamefont
  {J.~M.}\ \bibnamefont {Dahlstr{\"o}m}},\ }\href
  {https://doi.org/10.1103/PhysRevLett.123.133201} {\bibfield  {journal}
  {\bibinfo  {journal} {Physical Review Letters}\ }\textbf {\bibinfo {volume}
  {123}},\ \bibinfo {pages} {133201} (\bibinfo {year} {2019})}\BibitemShut
  {NoStop}%
\bibitem [{\citenamefont {De~Roulet}\ \emph {et~al.}(2024)\citenamefont
  {De~Roulet}, \citenamefont {Drescher}, \citenamefont {Sato},\ and\
  \citenamefont {Leone}}]{deroulet2024}%
  \BibitemOpen
  \bibfield  {author} {\bibinfo {author} {\bibfnamefont {B.~R.}\ \bibnamefont
  {De~Roulet}}, \bibinfo {author} {\bibfnamefont {L.}~\bibnamefont {Drescher}},
  \bibinfo {author} {\bibfnamefont {S.~A.}\ \bibnamefont {Sato}},\ and\
  \bibinfo {author} {\bibfnamefont {S.~R.}\ \bibnamefont {Leone}},\ }\href
  {https://doi.org/10.1103/PhysRevB.110.174301} {\bibfield  {journal} {\bibinfo
   {journal} {Physical Review B}\ }\textbf {\bibinfo {volume} {110}},\ \bibinfo
  {pages} {174301} (\bibinfo {year} {2024})}\BibitemShut {NoStop}%
\bibitem [{\citenamefont {Liao}\ \emph {et~al.}(2015)\citenamefont {Liao},
  \citenamefont {Sandhu}, \citenamefont {Camp}, \citenamefont {Schafer},\ and\
  \citenamefont {Gaarde}}]{liao2015}%
  \BibitemOpen
  \bibfield  {author} {\bibinfo {author} {\bibfnamefont {C.-T.}\ \bibnamefont
  {Liao}}, \bibinfo {author} {\bibfnamefont {A.}~\bibnamefont {Sandhu}},
  \bibinfo {author} {\bibfnamefont {S.}~\bibnamefont {Camp}}, \bibinfo {author}
  {\bibfnamefont {K.~J.}\ \bibnamefont {Schafer}},\ and\ \bibinfo {author}
  {\bibfnamefont {M.~B.}\ \bibnamefont {Gaarde}},\ }\href
  {https://doi.org/10.1103/PhysRevLett.114.143002} {\bibfield  {journal}
  {\bibinfo  {journal} {Physical Review Letters}\ }\textbf {\bibinfo {volume}
  {114}},\ \bibinfo {pages} {143002} (\bibinfo {year} {2015})}\BibitemShut
  {NoStop}%
\bibitem [{\citenamefont {Pfeiffer}\ \emph {et~al.}(2013)\citenamefont
  {Pfeiffer}, \citenamefont {Bell}, \citenamefont {Beck}, \citenamefont
  {Mashiko}, \citenamefont {Neumark},\ and\ \citenamefont
  {Leone}}]{pfeiffer2013}%
  \BibitemOpen
  \bibfield  {author} {\bibinfo {author} {\bibfnamefont {A.~N.}\ \bibnamefont
  {Pfeiffer}}, \bibinfo {author} {\bibfnamefont {M.~J.}\ \bibnamefont {Bell}},
  \bibinfo {author} {\bibfnamefont {A.~R.}\ \bibnamefont {Beck}}, \bibinfo
  {author} {\bibfnamefont {H.}~\bibnamefont {Mashiko}}, \bibinfo {author}
  {\bibfnamefont {D.~M.}\ \bibnamefont {Neumark}},\ and\ \bibinfo {author}
  {\bibfnamefont {S.~R.}\ \bibnamefont {Leone}},\ }\href
  {https://doi.org/10.1103/PhysRevA.88.051402} {\bibfield  {journal} {\bibinfo
  {journal} {Physical Review A}\ }\textbf {\bibinfo {volume} {88}},\ \bibinfo
  {pages} {051402} (\bibinfo {year} {2013})}\BibitemShut {NoStop}%
\bibitem [{\citenamefont {Gaynor}\ \emph {et~al.}(2021)\citenamefont {Gaynor},
  \citenamefont {Fidler}, \citenamefont {Lin}, \citenamefont {Chang},
  \citenamefont {Zuerch}, \citenamefont {Neumark},\ and\ \citenamefont
  {Leone}}]{gaynor2021}%
  \BibitemOpen
  \bibfield  {author} {\bibinfo {author} {\bibfnamefont {J.~D.}\ \bibnamefont
  {Gaynor}}, \bibinfo {author} {\bibfnamefont {A.~P.}\ \bibnamefont {Fidler}},
  \bibinfo {author} {\bibfnamefont {Y.-C.}\ \bibnamefont {Lin}}, \bibinfo
  {author} {\bibfnamefont {H.-T.}\ \bibnamefont {Chang}}, \bibinfo {author}
  {\bibfnamefont {M.}~\bibnamefont {Zuerch}}, \bibinfo {author} {\bibfnamefont
  {D.~M.}\ \bibnamefont {Neumark}},\ and\ \bibinfo {author} {\bibfnamefont
  {S.~R.}\ \bibnamefont {Leone}},\ }\href
  {https://doi.org/10.1103/PhysRevB.103.245140} {\bibfield  {journal} {\bibinfo
   {journal} {Physical Review B}\ }\textbf {\bibinfo {volume} {103}},\ \bibinfo
  {pages} {245140} (\bibinfo {year} {2021})}\BibitemShut {NoStop}%
\bibitem [{\citenamefont {Gaynor}\ \emph {et~al.}(2023)\citenamefont {Gaynor},
  \citenamefont {Fidler}, \citenamefont {Kobayashi}, \citenamefont {Lin},
  \citenamefont {Keenan}, \citenamefont {Neumark},\ and\ \citenamefont
  {Leone}}]{gaynor2023}%
  \BibitemOpen
  \bibfield  {author} {\bibinfo {author} {\bibfnamefont {J.~D.}\ \bibnamefont
  {Gaynor}}, \bibinfo {author} {\bibfnamefont {A.~P.}\ \bibnamefont {Fidler}},
  \bibinfo {author} {\bibfnamefont {Y.}~\bibnamefont {Kobayashi}}, \bibinfo
  {author} {\bibfnamefont {Y.-C.}\ \bibnamefont {Lin}}, \bibinfo {author}
  {\bibfnamefont {C.~L.}\ \bibnamefont {Keenan}}, \bibinfo {author}
  {\bibfnamefont {D.~M.}\ \bibnamefont {Neumark}},\ and\ \bibinfo {author}
  {\bibfnamefont {S.~R.}\ \bibnamefont {Leone}},\ }\href
  {https://doi.org/10.1103/PhysRevA.107.023526} {\bibfield  {journal} {\bibinfo
   {journal} {Physical Review A}\ }\textbf {\bibinfo {volume} {107}},\ \bibinfo
  {pages} {023526} (\bibinfo {year} {2023})}\BibitemShut {NoStop}%
\bibitem [{\citenamefont {Drescher}\ \emph {et~al.}(2020)\citenamefont
  {Drescher}, \citenamefont {Vrakking},\ and\ \citenamefont
  {Mikosch}}]{drescher2020}%
  \BibitemOpen
  \bibfield  {author} {\bibinfo {author} {\bibfnamefont {L.}~\bibnamefont
  {Drescher}}, \bibinfo {author} {\bibfnamefont {M.~J.~J.}\ \bibnamefont
  {Vrakking}},\ and\ \bibinfo {author} {\bibfnamefont {J.}~\bibnamefont
  {Mikosch}},\ }\href {https://doi.org/10.1088/1361-6455/ab9765} {\bibfield
  {journal} {\bibinfo  {journal} {Journal of Physics B: Atomic, Molecular and
  Optical Physics}\ }\textbf {\bibinfo {volume} {53}},\ \bibinfo {pages}
  {164005} (\bibinfo {year} {2020})}\BibitemShut {NoStop}%
\bibitem [{\citenamefont {Geondzhian}\ \emph {et~al.}(2022)\citenamefont
  {Geondzhian}, \citenamefont {Rubio},\ and\ \citenamefont
  {Altarelli}}]{geondzhian2022}%
  \BibitemOpen
  \bibfield  {author} {\bibinfo {author} {\bibfnamefont {A.}~\bibnamefont
  {Geondzhian}}, \bibinfo {author} {\bibfnamefont {A.}~\bibnamefont {Rubio}},\
  and\ \bibinfo {author} {\bibfnamefont {M.}~\bibnamefont {Altarelli}},\ }\href
  {https://doi.org/10.1103/PhysRevB.106.115433} {\bibfield  {journal} {\bibinfo
   {journal} {Physical Review B}\ }\textbf {\bibinfo {volume} {106}},\ \bibinfo
  {pages} {115433} (\bibinfo {year} {2022})}\BibitemShut {NoStop}%
\bibitem [{\citenamefont {Malakhov}\ \emph {et~al.}(2024)\citenamefont
  {Malakhov}, \citenamefont {Cistaro}, \citenamefont {Mart{\'i}n},\ and\
  \citenamefont {Pic{\'o}n}}]{malakhov2024}%
  \BibitemOpen
  \bibfield  {author} {\bibinfo {author} {\bibfnamefont {M.}~\bibnamefont
  {Malakhov}}, \bibinfo {author} {\bibfnamefont {G.}~\bibnamefont {Cistaro}},
  \bibinfo {author} {\bibfnamefont {F.}~\bibnamefont {Mart{\'i}n}},\ and\
  \bibinfo {author} {\bibfnamefont {A.}~\bibnamefont {Pic{\'o}n}},\ }\href
  {https://doi.org/10.1038/s42005-024-01689-4} {\bibfield  {journal} {\bibinfo
  {journal} {Communications Physics}\ }\textbf {\bibinfo {volume} {7}},\
  \bibinfo {pages} {1} (\bibinfo {year} {2024})}\BibitemShut {NoStop}%
\bibitem [{\citenamefont {{von Korff Schmising}}\ \emph
  {et~al.}(2017)\citenamefont {{von Korff Schmising}}, \citenamefont {Weder},
  \citenamefont {Noll}, \citenamefont {Pfau}, \citenamefont {Hennecke},
  \citenamefont {Str{\"u}ber}, \citenamefont {Radu}, \citenamefont {Schneider},
  \citenamefont {Staeck}, \citenamefont {G{\"u}nther}, \citenamefont
  {L{\"u}ning}, \citenamefont {el~dine Merhe}, \citenamefont {Buck},
  \citenamefont {Hartmann}, \citenamefont {Viefhaus}, \citenamefont {Treusch},\
  and\ \citenamefont {Eisebitt}}]{vonkorffschmising2017}%
  \BibitemOpen
  \bibfield  {author} {\bibinfo {author} {\bibfnamefont {C.}~\bibnamefont {{von
  Korff Schmising}}}, \bibinfo {author} {\bibfnamefont {D.}~\bibnamefont
  {Weder}}, \bibinfo {author} {\bibfnamefont {T.}~\bibnamefont {Noll}},
  \bibinfo {author} {\bibfnamefont {B.}~\bibnamefont {Pfau}}, \bibinfo {author}
  {\bibfnamefont {M.}~\bibnamefont {Hennecke}}, \bibinfo {author}
  {\bibfnamefont {C.}~\bibnamefont {Str{\"u}ber}}, \bibinfo {author}
  {\bibfnamefont {I.}~\bibnamefont {Radu}}, \bibinfo {author} {\bibfnamefont
  {M.}~\bibnamefont {Schneider}}, \bibinfo {author} {\bibfnamefont
  {S.}~\bibnamefont {Staeck}}, \bibinfo {author} {\bibfnamefont {C.~M.}\
  \bibnamefont {G{\"u}nther}}, \bibinfo {author} {\bibfnamefont
  {J.}~\bibnamefont {L{\"u}ning}}, \bibinfo {author} {\bibfnamefont
  {A.}~\bibnamefont {el~dine Merhe}}, \bibinfo {author} {\bibfnamefont
  {J.}~\bibnamefont {Buck}}, \bibinfo {author} {\bibfnamefont {G.}~\bibnamefont
  {Hartmann}}, \bibinfo {author} {\bibfnamefont {J.}~\bibnamefont {Viefhaus}},
  \bibinfo {author} {\bibfnamefont {R.}~\bibnamefont {Treusch}},\ and\ \bibinfo
  {author} {\bibfnamefont {S.}~\bibnamefont {Eisebitt}},\ }\href
  {https://doi.org/10.1063/1.4983056} {\bibfield  {journal} {\bibinfo
  {journal} {Review of Scientific Instruments}\ }\textbf {\bibinfo {volume}
  {88}},\ \bibinfo {pages} {053903} (\bibinfo {year} {2017})}\BibitemShut
  {NoStop}%
\bibitem [{\citenamefont {Yatsenko}\ \emph {et~al.}(1999)\citenamefont
  {Yatsenko}, \citenamefont {Halfmann}, \citenamefont {Shore},\ and\
  \citenamefont {Bergmann}}]{Yatsenko:1999aa}%
  \BibitemOpen
  \bibfield  {author} {\bibinfo {author} {\bibfnamefont {L.~P.}\ \bibnamefont
  {Yatsenko}}, \bibinfo {author} {\bibfnamefont {T.}~\bibnamefont {Halfmann}},
  \bibinfo {author} {\bibfnamefont {B.~W.}\ \bibnamefont {Shore}},\ and\
  \bibinfo {author} {\bibfnamefont {K.}~\bibnamefont {Bergmann}},\ }\href
  {https://doi.org/https://doi.org/10.1103/PhysRevA.59.2926} {\bibfield
  {journal} {\bibinfo  {journal} {Physical Review A}\ }\textbf {\bibinfo
  {volume} {59}},\ \bibinfo {pages} {2926} (\bibinfo {year}
  {1999})}\BibitemShut {NoStop}%
\bibitem [{\citenamefont {Kramida}\ \emph {et~al.}(2023)\citenamefont
  {Kramida}, \citenamefont {{Yu.~Ralchenko}}, \citenamefont {Reader},\ and\
  \citenamefont {{and NIST ASD Team}}}]{NIST_ASD}%
  \BibitemOpen
  \bibfield  {author} {\bibinfo {author} {\bibfnamefont {A.}~\bibnamefont
  {Kramida}}, \bibinfo {author} {\bibnamefont {{Yu.~Ralchenko}}}, \bibinfo
  {author} {\bibfnamefont {J.}~\bibnamefont {Reader}},\ and\ \bibinfo {author}
  {\bibnamefont {{and NIST ASD Team}}},\ }\href@noop {} {}\bibinfo
  {howpublished} {{NIST Atomic Spectra Database (ver. 5.11), [Online].
  Available: {\tt{https://physics.nist.gov/asd}} [2017, April 9]. National
  Institute of Standards and Technology, Gaithersburg, MD.}} (\bibinfo {year}
  {2023})\BibitemShut {NoStop}%
\bibitem [{\citenamefont {Delone}\ and\ \citenamefont
  {Krainov}(1999)}]{Delone:1999aa}%
  \BibitemOpen
  \bibfield  {author} {\bibinfo {author} {\bibfnamefont {N.~B.}\ \bibnamefont
  {Delone}}\ and\ \bibinfo {author} {\bibfnamefont {V.~P.}\ \bibnamefont
  {Krainov}},\ }\href {https://doi.org/10.1070/PU1999v042n07ABEH000557}
  {\bibfield  {journal} {\bibinfo  {journal} {Physics-Uspekhi}\ }\textbf
  {\bibinfo {volume} {42}},\ \bibinfo {pages} {669} (\bibinfo {year}
  {1999})}\BibitemShut {NoStop}%
\end{thebibliography}%

\end{document}